# MA-CDMR: An Intelligent Cross-domain Multicast Routing Method based on Multiagent Deep Reinforcement Learning in Multi-domain SDWN

Miao Ye, Hongwen Hu, Xiaoli Wang, *Member, IEEE,* Yuping Wang, *Senior Member, IEEE,* Yong Wang, Wen Peng, Jihao Zheng

*Abstract*—The cross-domain multicast routing problem in a software-defined wireless network (SDWN) with multiple controllers is a classic NP-hard combinatorial optimization problem. As the network size increases, designing and implementing cross-domain multicast routing paths in the network requires not only designing efficient solution algorithms to obtain the optimal cross-domain multicast tree but also ensuring the timely and flexible acquisition and maintenance of global network state information. However, existing solutions have a limited ability to sense the network traffic state, affecting the quality of service (QoS) of multicast services. In addition, these methods have difficulty adapting to the highly dynamically changing network states and have slow convergence speeds. To this end, this paper aims to design and implement a multiagent deep reinforcement learning (DRL)-based cross-domain multicast routing (MA-CDMR) method for SDWN with multicontroller domains. First, a multicontroller communication mechanism and a multicast group management module are designed to transfer and synchronize network information between different control domains of the SDWN, thus effectively managing the joining and classification of members in the cross-domain multicast group. Second, a theoretical analysis and proof show that the optimal cross-domain multicast tree includes an interdomain multicast tree and an intradomain multicast tree. An agent is established for each controller, and a cooperation mechanism between multiple agents is designed to effectively optimize cross-domain multicast routing and ensure consistency and validity in the representation of network state information for cross-domain multicast routing decisions. Third, a multiagent reinforcement learning-based method that combines online and offline training is designed to reduce the dependence on the real-time environment and increase the convergence speed of multiple agents. Finally, a series of experiments and their results show that the proposed method achieves good network performance under different network link information states, and the average bottleneck bandwidth is improved by 7.09%, 46.01%, 9.61%, and 10.11% compared with that of the KMB, SCTF, DRL-M4MR, and MADRL-MR method; the average delay is similar to that of MADRL-MR but significantly better than that of KMB, SCTF and DRL-M4MR, and the packet loss rate and average length of the multicast tree are also enhanced compared to those of the existing methods. The codes for DHRL-FNMRare open and available at https://github.com/GuetYe/DHRLFNMR

*Index Terms*—Enter keywords or phrases i

Manuscript received Aug ##, 2024. This work was supported in part by the National Natural Science Foundation of China (Nos.62161006, 62172095, 61861013), the subsidization of the Innovation Project of Guangxi Graduate Education (No. YCSW2022271), and the Guangxi Key Laboratory of Wireless Wideband Communication and Signal Processing (No. GXKL06220110). (Corresponding author: Xiaoli Wang).

Miao Ye, Hongwen Hu, Yong Wang, Jihao Zheng, and Wen Peng are with the School of Information and Communication, Guilin University of Electronic Technology, Guilin 541004, China(e-mail: ym@mail.xidian.ediu.cn, huhwown@foxmail.com, ywang@guet.edu.cn, wenpeng1994325@gmail.com, jihaozheng77@gmail.com ).
Xiaoli Wang and Yuping Wang are with the School of Computer Science and Technology at Xidian University, Xi'an, Shaanxi, 710071, China (e-mail: wangxiaoli@mail.xidian.edu.cn, ywang@xidian.edu.cn).

## I. INTRODUCTION

WITH the continuous development of wireless communication technology and the expansion of wireless network applications in various fields, wireless multicast communication, as an essential data transmission method, is widely used in multimedia conferencing, real-time video transmission, team collaboration, and distributed computing. Compared with point-to-point communication, multicast communication [1] can be employed to transmit the same data to multiple receivers to achieve efficient information distribution and sharing in wireless networks, efficiently reducing the network bandwidth and load. In essence, in a multicast routing method for multicast communication, an optimal multicast tree is built from the source node to all destination nodes [2]. The aim is to maximize the bottleneck bandwidth, minimize the transmission delay and packet loss rate, and improve network resource utilization. Constructing an optimal multicast tree in a high-speed and dynamically changing wireless network is an NP-hard combinatorial optimization problem. Therefore, there is a need to obtain as much information on global network link state states as possible in real time. However, traditional wireless network management methods could be more efficient, making it easier to obtain this information.

Software-defined wireless network (SDWN) technology [3] separates the network control plane from the data plane and uses centralized management and flexible programming mechanisms for network state information acquisition and global optimal network resource allocation. Therefore, this technology provides an excellentsolution to overcome the shortcomings of traditional wireless network management methods. However, as the network size and complexity increase, the scalability of the management mode of a single SDWN control domain becomes poor, and performance bottlenecks, such as challenges related to adapting to heterogeneity, are observed. To overcome these limitations and improve network performance and reliability, extending the SDWN from a single-controller domain to multicontroller domains is necessary. In an SDWN within a multicontroller domain, multicast issues include those related to message delivery and synchronization between controller domains, multicast group management, multicast routing table updating, and effectively constructing multicast routes across multiple controller domains.

In a multidomain wireless network, the traditional multiprotocol border gateway protocol (MBGP) [4] can transmit multicast routing information in a cross-domain environment by establishing and configuring peer-to-peer connections at the border router, and correct forwarding of the multicast data is realized on the basis of this information. However, using the MBGP for the configuration of interdomain routing in an SDWN with multiple controller domains is complex, and the lack of direct integration with controllers requires additional development and customization, which may lead to delays and inconsistencies in message delivery and synchronization [5]. In addition, since the intra- and interdomain network link information changes dynamically and multicast source nodes and multicast destination nodes may be distributed in different domains, the discovery of multicast groups, updating of multicast routing tables, construction of cross-domain multicast trees, and transmission of cross-domain data are challenging. Therefore, as part of the architecture of an SDWN with multiple controller domains, it is essential to design a cross-domain multicast tree that adapts to the dynamic changes in intra- and interdomain network link information to meet the needs of high-performance cross-domain multicast services.

To solve the NP-hard combinatorial optimization problem of finding the optimal multicast tree, common solutions include employing suitable optimal search methods, such as greedy search (e.g., the KMB [6] and SCTF [7] algorithms) and heuristic swarm intelligent optimization (such as the genetic algorithm (GA) [8]) algorithms. However, these algorithms lack flexibility and adaptability to high-speed and dynamically changing traffic demands.

Deep reinforcement learning (DRL) algorithms [9], which have been recently developed, are more flexible than common optimization algorithms and can adapt to complex network topologies and dynamically changing traffic demands. The data-driven learning ability of these algorithms enables them to extract features and patterns from large amounts of network data to generate an optimal multicast routing strategy, thereby improving the multicast routing performance. At present, deep reinforcement learning methods, such as Q-RL [10], DRL-M4MR [11], and MADRL-MR [12], which are reinforcement learning mechanism-based routing methods, are applied to solve multicast problems; however, these deep reinforcement learning-based methods are mostly limited to multicast problems in single-domain scenarios.

In addition, the current deep reinforcement learning methods for solving routing optimization problems are mostly online reinforcement learning methods [13], and the training strategies are divided into on-policy strategies [14] and off-policy strategies [15]. Both of these strategies require interaction with the environment. However, in a large-scale network environment, the cost of real-time interaction between an agent and the environment is high. The training process requires extensive trial and error and exploration. Most online reinforcement learning methods use only current real-time data, i.e., the utilization rate of historical data is low, which may result in high training costs and resource consumption. Because there is no environmental interaction in an offline reinforcement learning [16] method, such methods can be tailored to large-scale network environments, reducing the cost of real-time interactions with large-scale environments and improving the learning effect and training speed through the effective use of historical data. However, for a dynamically changing network environment in real time, the problems of sample bias and lack of exploration of offline data need to be solved.

To fully exploit the advantage of deep reinforcement learning in solving the multicast routing problem of the SDWN with multiple domains, a multiagent deep learning method based on an SDWN with multicontroller domains is applied, and a multiagent deep reinforcement learning-based intelligent cross-domain multicast routing (MA-CDMR) method is designed. To solve problems related to the synchronization and transfer of network information between domains and the discovery of multicast groups with group members distributed in different controller domains, a controller communication module and a multicast group management module are designed. These modules aim to maintain the consistency of information transfer and synchronization between different domains, identify cross-domain multicast groups and effectively manage the joining and leaving of multicast group members. A theoretical analysis and proof show that the optimal cross-domain multicast tree includes an interdomain multicast tree and an intradomain multicast tree. To solve the problem of constructing intradomain and interdomain multicast trees, some problem-solving agents are designed, and consistency and effectiveness in the representation of network state information are achieved when making cross-domain multicast routing decisions by designing collaborative mechanisms for the intelligent tasks that are relevant to the problem. To improve the training effect of multiple agents, accelerate the convergence speed, and reduce the cost of real-time interactions with large-scale environments, in this paper, offline reinforcement learning is employed to design a combined online and offline training method. In addition, a fully decentralized (FL) solution paradigm for multiple agents is designed to improve the cooperation efficiency of multiple agents.

The main research contributions of this paper are as follows:

1) In response to the characteristics of an SDWN with multiple domains, a framework for solving multicast problems on the basis of network state information awareness and multiagent deep reinforcement learning is proposed. In addition, a theoretical analysis of the multicast routing problem in the SDWN with multicontroller domains is presented, and the cross-domain multicast tree problem is broken down into two subproblems: interdomain multicast tree construction and intradomain multicast tree construction. Cooperative multiagent reinforcement learning algorithms are designed for these two subproblems.

2) Compared with the traditional MBGP, a controller communication mechanism and a multicast group management module are designed on the basis of the characteristics of the SDWN (such as control logic centralization and programmability). The aims are to achieve intercontroller domain message passing and synchronization and to identify and effectively manage cross-domain multicast group members. The designed controller communication and management mechanism can flexibly and conveniently acquire global network status information, realize the optimized distribution and utilization of multicast traffic resources, enhance coordination and cooperation between controller domains, and improve the overall performance and efficiency of the network.

3) A state space is designed to consider the cross-domain multicast issue in the SDWN with multicontroller domains. This state space includes matrices representing the bandwidth, delay, packet loss rate, packet error rate, distance between access points (APs), and multicast tree state of the network links. This design enables the agent to better perceive changes in the network link state information and the multicast tree construction process. Considering the interdomain and intradomain multicast issues, appropriate actions and strategies are designed, greatly improving the exploration efficiency of the agent. The designed action space of the agent for constructing the interdomain multicast tree is the set of all edges between the domains, and each action is selected as one of the edges. The designed action space of the agent for constructing the intradomain multicast tree is the set of nodes in the domain, and each action is selected as a hop node. Different reward and penalty functions are designed for different action strategies adopted by the agent, and they include a single-step decision reward, a path completion reward, an invalid action penalty, and a loop penalty. The agent is guided to construct efficient interdomain and intradomain multicast trees.

4) Multiple agents are designed to construct and optimize the cross-domain multicast tree through cooperative learning and policy coordination, and the FL solution paradigm is proposed to improve the stability of multiagent cooperation. A combined online and offline training method is designed to reduce the interaction frequency with the environment and the dependence on the real-time environment and to effectively increase the convergence speed of multiple agents.

The remainder of this paper is organized as follows: Section II describes related work. In Section III, the multicast, Steiner tree and cross-domain multicast tree problems are analyzed. Section IV describes the proposed MA-CDMR architecture. Section V describes the MA-CDMR algorithm in detail. Section VI describes the experimental environment and the performance evaluation results. Section VII presents conclusions and suggestions for future work.

## II. RELATED WORK

This section introduces technologies related to cross-domain multicast routing in an SDWN with multiple controller domains, and the advantages and disadvantages of related algorithms are analyzed.

Methods for traditional multicast tree problems: L. Kou et al.[6] proposed the KMB algorithm to construct a Steiner tree on the basis of the short path tree (SPT) and the minimum spanning tree (MST). Angelopoulos et al. [17] used the selective closest terminal first (SCTF) algorithm to construct a multicast tree, which calculates the paths from a source to all the destination nodes by employing Dijkstra's algorithm, and finally all the topological paths from the source to all the destination nodes are used as the approximate solution for the Steiner tree. Takahashi et al. [18] proposed the minimum path cost heuristic (MPH) algorithm. V. J. Rayward-Smith et al.[19] designed the average distance heuristic (ADH) algorithm. Naser et al. [20] proposed a mathematical model of a minimum-power multicast tree considering the residual power of each node in an SDN cluster environment and designed three clustering-based multicast routing strategies via the proposed model. Weixiao et al. [21] considered the network performance of multicast transmission in terms of bandwidth, utilization, and delay and proposed a multiconstraint multicast routing mutation mechanism. In addition, to address the high time complexity of the multicast tree generation algorithm, a multidomain multicontroller multicast routing algorithm was proposed, and it improved upon the SPT algorithm. The improved algorithm searches for the first k short paths, randomly selects all the paths from the multicast source to the destination for superposition, and uses a greedy algorithm to perform the unloop operation to obtain a multicast tree. Chiang et al. [22] discussed online distributed multicast traffic engineering for multidomain SDNs and designed a competitive distributed algorithm using the ideas of domain trees, dual candidate forest construction, and forest rerouting. The designed algorithm is used to collaboratively construct cross-domain multicast routes via controllers in each domain that share their network state information with controllers in neighboring domains and constructs a multicast tree via the SPT algorithm. Liu et al. [23] proposed a multiconstrained multiobjective path optimization algorithm (MCOPF) for the SD-WAN. The algorithm categorizes the request streams into delay-sensitive streams and other streams. In addition to guaranteeing quality of service (QoS) for each stream, the algorithm is designed with the goal of maximizing delay-sensitive stream requests and optimizing the utilization of network resources. To improve the effectiveness of the algorithm, a hierarchical multicontroller collaborative framework is used, which divides the control plane into multiple levels. The root controller manages the global view, and the domain controllers manage the regional networks. These traditional algorithms usually focus on a single optimization objective, such as minimizing the path length and bandwidth. However, a single objective does not accurately reflect the actual conditions of the network, and considering only one objective may lead to problems such as failure to achieve the global optimum and network congestion. Moreover, traditional algorithms are limited in scalability, and the exchange of routing information and the computational loads between the routers in a large-scale network could increase dramatically, resulting in the degradation of network stability and performance.

Intelligent Optimization Algorithms: Wu et al. [24] proposed a multicast restoration algorithm for multilayer multidomain optical networks on the basis of hybrid swarm intelligence. This algorithm combines the artificial fish model in the PCE-based optical network architecture with noncooperative game theory for intradomain path searches and employs an improved fruit fly optimization algorithm to find recovery paths. This approach enhances the convergence speed while achieving improved local and global search path selection. Ke et al. [8] introduced a multicast routing method based on genetic algorithms (GAs) designed to optimize multicast routing strategies for flow scheduling in large-data-center networks, considering both flow types and network states. Liu et al. [25] proposed a routing scheme based on a time-shift multilayer graph (TS-MLG) framework to consider both spatial and temporal factors and facilitate interdomain bulk data transfer. This scheme can be used to determine cross-domain routing paths, storage points, and transmission times through conventional routing calculations but only improves interdomain bandwidth utilization, with high network state maintenance costs and computational complexity. Zhao et al. [26] proposed an intelligent interdomain routing scheme supported by a hierarchical control plane structure based on subtopology graphs, considering differentiated QoS and energy savings. This scheme enables multidomain quality of transmission (QoT), energy-aware routing, and spectrum allocation (RSA), but it is only suitable for interdomain routing and does not address intradomain routing issues. Li et al. [27] designed a scalable and protocol-independent path algorithm for multidomain packet networks, and it is based on deep learning, which supports protocol-independent forwarding in the data plane. However, this algorithm only considers bandwidth as a QoS requirement, neglecting information such as the link delay and packet loss rate, and its high training complexity makes it challenging to adapt to dynamically changing network environments.

Reinforcement learning-based methods: Xu et al. [28] proposed a hierarchical reinforcement learning framework for multidomain elastic optical networks to realize RMSA)= for interdomain service requests. The framework consists of a high-level DRL module and multiple low-level DRL modules (one for each domain), with the DRL modules collaborating with each other. For interdomain service requests, the high-level module obtains some abstract information from the low-level DRL modules and generates an interdomain RMSA decision for low-level modules. However, the method is only concerned with interdomain service request routing. Dang et al. [29] proposed an efficient link-state multicast routing method by optimizing link weights with multiagent reinforcement learning (MARL). The method first provides an integer linear programming (ILP) formula to find the optimal link weights for SPT-based multicast routing protocols to minimize the total network cost and then designs a MARL solution to optimize the link weight problem for efficient multicast routing in a distributed manner. However, the proposed MARL method is only used to optimize the link weights and does not consider the interactions among multiple intelligent agents. Zhao et al. [30] proposed a MARL-based cross-domain SFC routing method. The method combines the embedding costs associated with the paths, delays and loads for all domains and uses the resulting value as the construction cost of routing. However, the method only considers the network chain delay and does not fully consider other types of network chain information, such as the bandwidth and packet loss rate. Bhavanasi et al. [31] proposed a resilient routing method based on graph neural networks and MARL. The method uses a graph neural network algorithm to design reinforcement learning intelligent agents, and a network traffic dataset encoded in a graph format is generated as the training input. In this method, the reinforcement learning strategy (i.e., a given reward function) can be applied to any topology and learn from changes in routing and congestion events without the need to retrain the neural network. However, the method's ability to relearn the optimal strategy in cases with large networks is limited. Ye et al. [32] proposed a cross-domain intelligent routing algorithm for SDNs that is based on MARL and network traffic prediction. In this method, a Dueling DQN intelligent agent is established for each domain, and recurrent neural networks are used for network traffic prediction. The actions of the intelligent agents are designed as $k$-paths, and all k paths are generated via Dijkstra's algorithm, allowing the agent to learn to select the optimal path. However, the generated k paths are fixed and cannot adapt to dynamic changes in network link traffic, resulting in local optimum problems. Casas-Velasco et al. [10] proposed a Q-learning based multicast method that considers link state information for routing decisions, focusing mainly on unicast routing. C. Zhao et al. [11] developed a deep reinforcement learning-based intelligent multicast routing method in SDN, which considers link bandwidth, delay, and packet loss rate. However, this method suffers from slow agent convergence.

To solve the above problems, in this paper, the MA-CDMR method in an SDWN with multicontroller domains is designed and implemented. A multicontroller communication mechanism and a multicast group management module are designed to transfer and synchronize network information between different control domains of the SDWN, and the optimal cross-domain multicast tree is decomposed into two parts—the interdomain multicast tree and the intradomain multicast tree. In addition, a multiagent reinforcement learning-based solution algorithm is designed for each controller. Finally, a multiagent reinforcement learning-based method that combines online and offline training is designed to reduce the dependence on the real-time environment and accelerate the convergence speed of multiple agents.

## III. Description and Modeling of the Cross-Domain Multicasting Problem

### A. Multicasting and the Steiner tree problem

Multicasting is a communication method in which data are simultaneously transmitted to multiple target nodes in a computer network. A multicasting method aims to construct a multicast tree so that the transmission cost is minimized. A multicast tree is a tree structure with a multicast source as the root node and covers all destination nodes. By constructing a multicast tree with minimum cost, data can be effectively transmitted to all destination nodes while reducing the network bandwidth and transmission delay. The minimum cost multicast tree problem actually corresponds to a problem in graph theory, namely, the minimum Steiner tree problem [33].

For a given wireless weighted graph $G(V, E, w)$, $V$ is the set of nodes in graph $G$, $E$ is the set of edges in graph $G$, $w$ is the edge weight, $e_{ij}$ represents an edge from node $i$ to node j, $e_{ij} \in E$, and the edge weight set is $w(e_{ij})$. For a set of multicast nodes $M \subseteq V$ and $M = \{\text{src}\} \cup D$, src is the source node and $D$ is the set of multicast destination nodes, where $D = \{d_1, d_2, \cdots, d_n\}$. Graph $G'$ is the subgraph of graph $G$ containing the set of multicast nodes $M$. Graph $G'$ also contains some nodes that are not in $M$, which are referred to as Steiner nodes.

In the minimum Steiner tree problem, the aim is to find a spanning tree $T = (V_T, E_T)$ containing $M$ in the graph $G'$ that can minimize the sum of the weights of the edges, as shown in (1).

$$\min_{T \subseteq G, M \subseteq V_T} \sum_{e_{ij} \in E_T} w(e_{ij}) \quad (1)$$

where $V_T$ represents all nodes in the tree $T$ and $E_T$ represents all edges in the tree $T$.

### B. Cross-domain multicast tree problem

This paper investigates the cross-domain multicast problem and introduces a solution method for the proposed SDWN with multicontroller domains. It is assumed that the subsequent analysis is performed based on a multidomain scenario; that is, the entire wireless network can be reasonably divided into $m$ control domains, denoted as $N_i, i = 1, \ldots, m$. Notably, dividing the control domains of the SDWN is a crucial process. In the subsequent discussion of cross-domain multicasting, a definitive division of the entire network and the dynamic migration of software-defined networking (SDN) switches within each domain are assumed. In addition, it is further assumed that the interdomain path cost between any two domains is greater than the path cost between any two nodes in the domain. On the basis of the above description of the problem background, the subsequent research in this paper is performed under the premise that the following two hypotheses are valid.

**Hypothesis 1**: After the rational division of the SDWN, the SDN switch nodes and their interconnections within each domain are fixed, eliminating the need to consider node movements.

**Hypothesis 2**: The interdomain path cost associated with the link between any two domains is greater than the cost for the path between any two nodes in the domain.

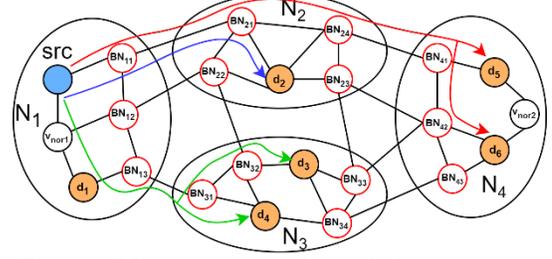

Figure 1 Multicast tree in the multidomain scenario

The multicast tree includes the multicast source node and the multicast destination nodes. Section III describes the scenario in which all these nodes are in the same domain. In this section, the scenario in which these nodes are in different domains is discussed. The multicast source is $src$, and the destination node of multicasting is $d_i$, where $i = 1, \ldots, n$. Some of the $n$ multicast destination nodes may be in the same control domain as $src$, and some may not be in the same control domain. The multicast routing path includes an intradomain and interdomain routing paths. As shown in Fig. 1, the blue node is $src$, the orange node is $d_i$, and the red nodes are the boundary nodes (BNs) between the domains, which are not general nodes; therefore, $BN_{i,a}$ represents the $a$ th boundary node in domain $N_i$, the mapping function $dm: N_i \to \{d_j\}$ represents a set containing the destination node $d_j$ in any given domain $N_i$, the mapping function $Domain(d_i): d_i \to N_j$ represents the corresponding domain $N_j$ for any given destination node $d_i$, and the mapping function $BND: N_i \to \{BN_{ia}\}$ represents a set of boundary nodes $BN_{ia}$ corresponding to any given domain $N_i$. The three mapping functions $dm(N_i)$, $Domain(d_i)$ and $BND(N_i)$ are determined after the division of multiple controller domains, i.e., they are known.

The multicast routing path includes two parts: the intradomain routing path (for convenience of expression, this path is also referred to as the intradomain multicast tree in the subsequent parts of this paper) and the interdomain routing path (for convenience of expression, this path is referred to as the interdomain multicast tree in the subsequent parts of this paper). An optimal cross-domain multicast tree can be obtained based on the following definitions.

**Definition 1.** In a multicast tree, from the source node $src$ to the destination node $d_i$, the interdomain path $PN_i$ encompasses the following components from the source node $src$ to the destination node $d_i$: the domain sequence of each passed control domain $PN_i = <N_{i,1}, N_{i,2}, \ldots, N_{i,j}, \ldots, N_{i,|PN_i|}>$ $= (\mathcal{N}_i, \mathcal{E}_i)$, where $N_{i,j} \in \mathcal{N}_i$ represents the $j$th passed domain along the interdomain path $PN_i$; $\mathcal{N}_i$, which is the set of all passed domains along interdomain path $PN_i$; and $\mathcal{E}_i$, which is the connectivity relationship between the domains in $\mathcal{N}_i$, that is, the edge $<N_{i,j-1}, N_{i,j}>$ between adjacent domains $N_{i,j-1}$ and $N_{i,j}$ along interdomain path $PN_i$.

For example, in Fig. 1, the three interdomain paths represented by red, blue and green are $PN_2 = <N_1, N_2>$, $PN_3 = PN_4 = <N_1, N_3>$, and $PN_5 = PN_6 = <N_1, N_2, N_4>$, respectively.

**Definition 2** Interdomain multicast tree: A tree corresponds to the interdomain path $PN_i$, where $i = 1, \ldots, n$ from the source node $src$ to each destination node $d_i$. This tree only contains the links between the domains and can be considered a multicast tree in a graph formed by using the domains as "nodes" and the adjacent connectivity relationships between domains as "edges."

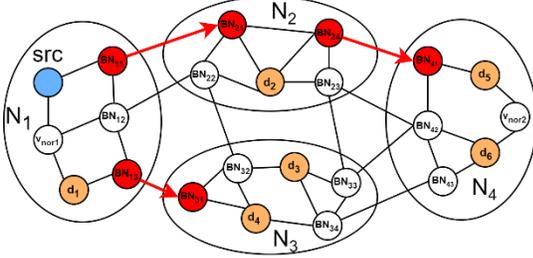

Figure 2 Interdomain multicast tree

The n interdomain paths $PN_i$, where $i = 1, \ldots, n$, from the source node $src$ to destination node $d_i$ can be used to represent an arbitrary interdomain multicast tree $T_{int}$, that is, $T_{int} = \{PN_1, \ldots, PN_i, \ldots PN_n\}$. The interdomain multicast tree $T_{int}$ can also be represented by the BN edges between domains, with edge set $P_{int} = \{(BN_{ia}, BN_{jb}) | BN_{ia} \in BND(N_i), BN_{ja} \in BND(N_j)\}$, where $(BN_{ia}, BN_{jb})$ represents the edge connecting the domains $N_i$ and $N_j$, $BN_{ia}$ represents the $a$th BN in domain $N_i$, and $BN_{ja}$ represents the $b$th BN in domain $N_j$. In this case, $T_{int} = P_{int}$. An interdomain multicast tree is shown in Fig. 2, where $BN_{11}, BN_{13}, BN_{21}, BN_{24}, BN_{34}$, and $BN_{43}$ are the selected BNs, $N_1$ is the source domain, $N_2, N_3$, and $N_4$ are the destination domains, and the interdomain multicast tree can be expressed as $P_{int} = \{(BN_{11}, BN_{21}), (BN_{13}, BN_{21}), (BN_{24}, BN_{41}), (BN_{34}, BN_{43})\}$. In general,

$$T_{int} = \{PN_1, \cdots, PN_i, \cdots PN_n\} \quad \text{or}$$
$$T_{int} = P_{int} = \{(BN_{ia}, BN_{jb}) | BN_{ia} \in BND(N_i), BN_{jb} \in BND(N_j)\} \quad (2)$$

The defined interdomain multicast tree in **Definition 2** is actually the interdomain routing path component of a cross-domain multicast tree. For the intradomain routing path of the cross-domain multicast tree, the following definitions apply to an intradomain multicast tree and an intradomain multicast forest.

**Definition 3** The intradomain multicast tree and intradomain multicast forest $T_i$ of domain $N_i$: From the source node $src$ to each destination node $d_i$, where $i = 1, \ldots, n$, for the intradomain routing path in each domain, $T_i$ is used to represent the path within domain $N_i$. Obviously, as long as a path passes through one domain $N_i$, then $T_i \neq \emptyset$. $T_i$ may be a single tree or a forest consisting of multiple trees, which can be defined as follows.

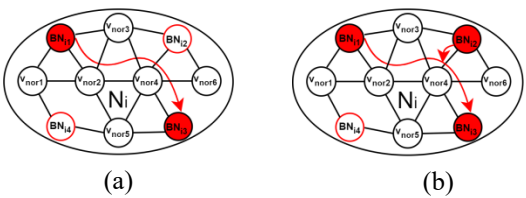

(a)      (b)

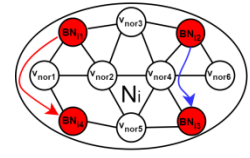

(c)
Figure 3 No destination node in the domain

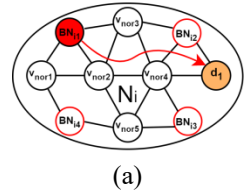 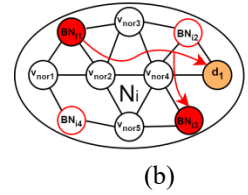

(a)      (b)

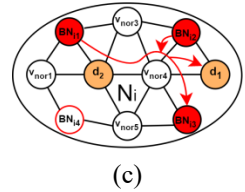 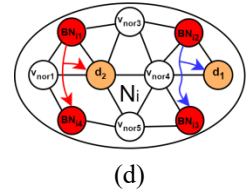

(c)      (d)
Figure 4 Multiple destination nodes in the domain

(1) There is no destination node in domain $N_i$ for any multicast tree. In this case, the multicast path only passes through this domain, i.e., enters domain $N_i$ from $BN_{ia}$ and leaves domain $N_i$ from $BN_{ib}$. If there is only one $BN_{ia}$ for entry into domain $N_i$, then the path in domain $N_i$ has only one connected component. At this time, $T_i$ can be considered a tree with $BN_{ia}$ as the root and $BN_{ib}$ as the leaf node, which is referred to as an intradomain multicast tree $T_i$, as shown in Fig. 3(a). If there are multiple $BN_{ia}$s for entry into domain $N_i$, then the path in domain $N_i$ may have a connected component, as shown in Fig. 3(b), or multiple connected components, as shown in Fig. 3(c). At this time, $T_i$ can be considered a forest consisting of trees with multiple $BN_{ia}$s as the roots, which is referred to as an intradomain multicast forest $T_i$.

(2) There is a destination node $d_i \in dm(N_i)$ in domain $N_i$ to be reached in the multicast tree. If there is only one such destination node, that is, $|dm(N_i)| = 1$, the multicast path only enters the domain $N_i$ via a unique $BN_{ia}$. At this time, $T_i$ is a tree with $BN_{ia}$ as the root (if there is no BN for leaving domain $N_i$, then the tree $T_i$ is a chain, as shown in Fig. 4(a); if there is $BN_{ib}$ for leaving domain $N_i$ to enter another domain, then the tree $T_i$ is a tree in the ordinary sense with $d_i$ and $BN_{ib}$ as leaf nodes and is referred to as an intradomain multicast tree $T_i$, as shown in Fig. 4(b)); if there are multiple destination nodes, that is, $|dm(N_i)| > 1$, then the multicast path can enter domain $N_i$ through multiple $BN_{ia}$s to reach these destination nodes, and the path in the domain $N_i$ may have a connected component, as shown in Fig. 4(c), or multiple connected components, as shown in Fig. 4(d). At this time, $T_i$ can be considered a forest consisting of trees with multiple $BN_{ia}$s as roots and is referred to as an intradomain multicast forest $T_i$.

The intradomain multicast tree and intradomain multicast forest $T_i$ in domain $N_i$ defined in Definition 3 can be simply represented as an edge set.

$$T_i = \{(v_1, v_2), (v_2, v_3), \cdots (v_{i,l}, v_{i,l-1})\} \quad (3)$$

where $(v_{i,l-1}, v_{i,l})$ represents the edge connecting node $v_{i,l-1}$ and node $v_{i,l}$ in domain $N_i$.

For a domain $N_i$ that has no interdomain multicast tree passing through it, an intradomain multicast forest $T_i \neq \emptyset$ exists. The set $T_{intra}$ of multicast trees of all domains $N_i (i = 1, \ldots, m)$ is expressed as:

$$T_{intra} = \bigcup_{i=1}^{m} T_i \quad (4)$$

where $m$ represents the number of domains.

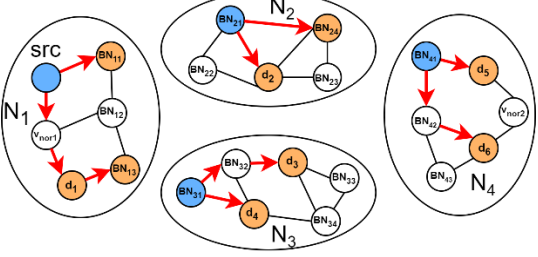

Figure 5 Example of an intradomain

Fig. 5 shows an example of an intradomain multicast tree. The four intradomain multicast trees are represented by $T_1, T_2, T_3$ and $T_4$. Fig. 5 shows that $src$ is the source node, $d_1, d_2, d_3, d_4, d_5$ and $d_6$ are the destination nodes, $v_{nor1}$ and $v_{nor2}$ are common nodes, $T_1 = \{(src, BN_{11}), (src, v_{nor1}), (v_{nor1}, d_1), (d_1, BN_{13})\}$, $T_2 = \{(BN_{21}, BN_{24}), (BN_{21}, d_2)\}$, $T_3 = \{(BN_{31}, BN_{32}), (BN_{32}, d_3), (BN_{31}, d_4)\}$, and $T_4 = \{(BN_{41}, d_5), (BN_{41}, BN_{42}), (BN_{42}, d_6)\}$.

**Definition 4** Cross-domain multicast tree $T_{cd}$: All the paths associated with $T_{int}$ and $T_{intra}$ are combined to form a cross-domain multicast tree $T_{cd}$,

$$T_{cd} = T_{int} \cup T_{intra} \quad (5)$$

**Definition 5** Optimal cross-domain multicast tree: Each multicast tree can be defined as a set of paths $p_i (i = 1, \ldots, n)$ from the source node $src$ to $n$ destination nodes $d_i (i = 1, \ldots, n)$ (similar to the definition of the multicast tree in our previous work MADRL-MR [12]) and can be expressed as $T = \{p_1, \cdots, p_k, \cdots, p_n\}$, where $p_k$ is the path from source node $src$ to destination node $d_k$ in multicast tree $T$ and where $n$ is the number of destination nodes. If each path $p_k \in T$ has a minimum cost, then the multicast tree $T$ is an end-to-end minimum cost tree. The cost of each path $p_k$ is defined as $c_k$. The optimization objectives are to maximize the bottleneck bandwidth $bw_k$; minimize the delay $delay_k$, packet loss rate $loss_k$, and packet error rate $err_k$; and calculate the distance to the wireless access point AP $dist_k$. All the indicator parameters are normalized to $[0,1]$ via max–min normalization [34]. For each path $p_k$, the construction cost can be expressed as $c_k$:

$$\cos t(p_k) = c_k = \beta_1 (1 - bw_k) + \beta_2 delay_k \\ + \beta_3 loss_k + \beta_4 err_k + \beta_5 dist_k \quad (6)$$

where:

$$bw_k = \min_{e_{ij} \in p_k} (bw_{ij})$$

$$delay_k = \sum_{e_{ij} \in p_k} delay_{ij}$$

$$loss_k = 1 - \prod_{e_{ij} \in p_k} (1 - loss_{ij}) \quad (7)$$

$$err_k = 1 - \prod_{e_{ij} \in p_k} (1 - err_{ij})$$

$$dist_k = average\left(\sum_{e_{ij} \in p_k} dist_{ij}\right)$$

where $p_k$ is the path from the source node $src$ to $d_k$ in multicast tree $T$, $bw_{ij}$ represents the remaining bandwidth of $e_{ij}$, $delay_{ij}$ represents the delay of $e_{ij}$, $loss_{ij}$ represents the packet loss rate of $e_{ij}$, $err_{ij}$ is the packet error rate of $e_{ij}$, and $dist_{ij}$ represents the distance of $e_{ij}$.

The optimal cross-domain multicast tree means to search for a Steiner tree $T_{cd}^*$ where the sum of path costs $c_k$ from the source node to all the destination nodes in this Steiner tree is minimized, as shown in Eq. (8).

$$T_{cd}^* = \underset{T_{cd}}{\arg\min} \cos t(T_{cd}) \\ = \underset{p_k \in T_{cd}}{\arg\min} \sum_{k=1}^{n} \cos t(p_k) \quad (8)$$

The least-cost multicast tree $T$ is the optimal cross-domain multicast tree. Based on Eq. (5), the following equation can be established:

$$\cos t(T_{cd}) = \cos t(T_{int} \cup T_{intra}) \quad (9)$$

By using Definitions 1 to Definitions 5, the optimal cross-domain multicast tree is defined. On the basis of these definitions, we know that the cross-domain multicast tree can be decomposed into interdomain and intradomain multicast trees to obtain solutions. On the basis of Hypothesis 2, "Assuming the path cost between any two domains is greater than the path cost between any two nodes within the domain", the following properties of the optimal cross-domain multicast tree can be obtained. The optimal multicast tree routing path does not include the interdomain path loop. On the basis of Fig. 6, this property is analyzed, and the related theorems and mathematical proofs are presented.

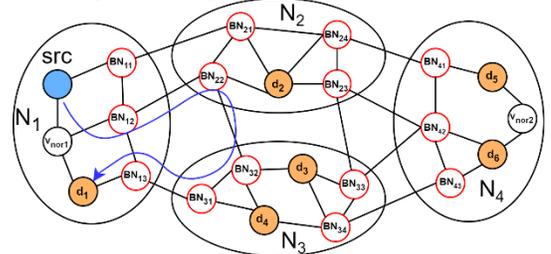

(a) Source node and destination nodes are in the same domain

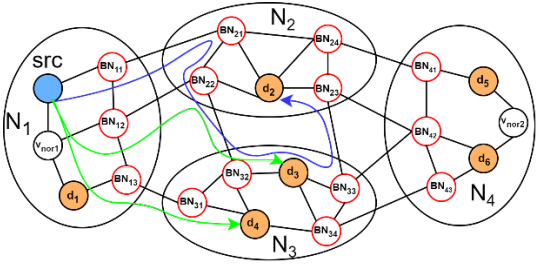

(b) Source node and destination nodes are in different domains

Figure 6 In a multidomain scenario, the source node and the destination node are in the same domain or in different domains

The situations (from a source node to any destination node) in Fig. 6 are analyzed as follows:

(1) When the multicast source node and any one of the multicast destination nodes are in the same domain, as shown in Fig. 6(a), $src$ and $d_1$ are in the same domain $N_1$, and the path from $src$ to $d_1$ is shown by the blue route in Fig. 6(a), i.e., starting from $N_1$, passing through the domains $N_2$ and $N_3$ and returning to $N_1$, thus forming an interdomain path loop $N_1 \to N_2 \to N_3 \to N_1$. In such a situation, the loop path is usually long, which involves the cooperative operation of controllers in different domains and leads to additional multidomain costs. On the basis of Hypothesis 2, the cost of the interdomain path loop $N_1 \to N_2 \to N_3 \to N_1$ is greater than that of the path of reaching the destination node from the source node $src$ in domain $N_1$ directly through the switchers in domain $N_1$. Therefore, a cross-domain path from a source node to a destination node in the same domain does not exist; i.e., nodes on the path from $src$ to $d_1$ are all in the domain $N_1$.

(2) A situation exists in which the source node and the destination node are not in the same domain, as shown in Fig. 6(b), where $src$ is in domain $N_1$, $d_2$ is in domain $N_2$, and $d_3$ and $d_4$ are in domain $N_3$, as shown in Fig. 6(b). The blue path starts from $N_1$, passes through $N_2$ and $N_3$ and returns back to $N_2$ to destination node $d_2$, thus forming an interdomain path loop $N_1 \to N_2 \to N_3 \to N_2$. Similar to the conclusions based on Fig. 6(a) based on Hypothesis 2, after arriving at domain $N_2$ from $src$, a path in domain $N_2$ cannot 'return' to domain $N_2$. Therefore, there is no interdomain path loop for the path from $src$ to the destination node $d_2$.

(3) A situation exists in which a source node and multiple destination nodes are in the same domain, as shown by the green path in Fig. 6(b). In this figure, destination nodes $d_3$ and $d_4$ are in the same domain $N_3$. The path is from $src$ to $d_3$ through $BN_{12}$, $N_3$ is entered through $BN_{22}$ in $N_2$, and the path from $src$ to $d_4$ involves directly entering $N_3$ through $BN_{13}$. Therefore, there are two different cross-domain paths between $N_1$ and $N_3$, namely, $N_1 \to N_3$ and $N_1 \to N_2 \to N_3$. If the costs of these two paths are $cost(N_1 \to N_3 \to d_3)$ and $cost(N_1 \to N_2 \to N_3 \to d_4)$, then

$cost(N_1 \to N_3 \to d_3) + cost(N_1 \to N_2 \to N_3 \to d_4)$
$= cost(N_1 \to N_3) + cost(BN_{32} \to d_3)$
$\qquad + cost(N_1 \to N_2 \to N_3) + cost(BN_{31} \to d_4)$
$> 2 \times \min(cost(N_1 \to N_3), cost(N_1 \to N_2 \to N_3))$
$\qquad + cost(BN_{32} \to d_3) + cost(BN_{31} \to d_4)$

The "greater than" denotation is also based on Hypothesis 2. As such, along the route from a source node to multiple different destination nodes in the same domain, one cannot travel via different interdomain paths.

To summarize the analysis of these situations, the following theorems are established.

**Theorem 1:** For the path of the optimal cross-domain multicast tree $T_{cd} = T_{int} \cup T_{intra}$,

(1) For the interdomain path sequence from source node $src$ to destination node $d_i$,
$PN_i =< N_{i,1}, N_{i,2}, \dots, N_{i,j}, \dots, N_{i,k}, \dots, N_{i,|PN_i|} >$,
and when $\forall j, k$ and $j \neq k$, $N_{i,j} \neq N_{i,k}$ is valid.

(2) For any two destination nodes $d_i$ and $d_j$, when $\text{Domain}(d_i) = \text{Domain}(d_j)$, $PN_i = PN_j$ is valid.

**Proof:** (1) This theorem indicates that an interdomain path sequence $PN_i$ that enters domain $N_j$ does not return to domain $N_j$ after passing through other domains. We can prove this theorem via the proof by contradiction. Suppose that Theorem (1) is not valid. If $j$ and $k$ exist, when $j \neq k$, $N_{i,j} = N_{i,k}$, and $\text{cost}(PN_i) = \text{cost}(N_{i,1} \to N_{i,j}) + \text{cost}(N_{i,j} \to N_{i,k}) + \text{cost}(N_{i,k} \to N_{i,j}) > \text{cost}(N_{i,1} \to N_{i,j}) + \text{cost}(N_{i,k} \to N_{i,j})$, which shows that an interdomain path sequence to the destination node $d_i$ that minimizes cost exists, contradicting the premise that $T_{cd}$ is the optimal cross-domain multicast tree; thus, the original proposition (1) holds.

(2) This conclusion means that the route from a source node to different destination nodes in the same domain does not require different interdomain paths. $PN_i =< N_{i,1}, N_{i,2}, \dots, N_{i,k}, \dots, N_{i,|PN_i|} >$ and $PN_j =< N_{j,1}, N_{j,2}, \dots, N_{j,k}, \dots, N_{i,|PN_i|} >$ reach destination nodes $d_i$ and $d_j$, respectively. Because $\text{Domain}(d_i) = \text{Domain}(d_j)$, these two destination nodes are in the same domain $N_h$. $PN_i$ enters domain $N_h$ through $BN_{h1}$ of $N_h$ and arrives at $d_i$, and $PN_j$ enters domain $N_h$ through $BN_{h2}$ of $N_h$ and arrives at $d_j$. The total cost of these two paths is

$\text{cost}(PN_i) + \text{cost}(PN_j)$
$= \text{cost}(PN_i) + \text{cost}(BN_{h1} \to d_i) + \text{cost}(PN_j)$
$\qquad + \text{cost}(BN_{h2} \to d_j)$

If $PN_i$ is the path with the lowest cost among the two paths of $PN_i$ and $PN_j$, on the basis of Hypothesis 2, for the two nodes $BN_{h1}$ and $d_j$ in domain $N_h$, $\text{cost}(BN_{h1} \to d_j)$ is less than any interdomain path. Therefore, we have

$\text{cost}(PN_i) + \text{cost}(PN_j)$
$= \text{cost}(PN_i) + \text{cost}(BN_{h1} \to d_i) + \text{cost}(PN_j)$
$\qquad + \text{cost}(BN_{h2} \to d_j)$
$\geq \text{cost}(PN_i) + \text{cost}(BN_{h1} \to d_i)$
$\qquad + \text{cost}(BN_{h1} \to d_j)$

Only when $PN_i$ and $PN_j$ are the same path, that is, $PN_i = PN_j$, would $\text{cost}(PN_i)$ be calculated only once. Therefore, if and only if $PN_i = PN_j$, the equality is valid.

Starting from the source node, one can first travel the path with the lowest cost among the two paths $PN_i$ and $PN_j$ (e.g., $PN_i$) to enter domain $N_h$ through $BN_{h1}$ of $N_h$ and then arrive at the destination nodes $d_i$ and $d_j$ separately. Therefore, for a cross-domain multicast tree $T_{cd}$, when $\text{Domain}(d_i) =$

Domain($d_j$), only when $PN_i = PN_j$ is $T_{cd}$ the optimal cross-domain multicast tree. Therefore, the original proposition (2) holds.

**Theorem 2:** For any domain through which the path of the optimal cross-domain multicast tree $T_{cd} = T_{int} \cup T_{intra}$ passes, only an intradomain multicast tree exists. However, an intradomain multicast forest does not exist.

The conclusion and proof of Theorem 1 show that there is only one BN entering any domain, which means that only one tree exists, and multiple trees never exist. Therefore, an intradomain multicast forest does not exist. Owing to the length limitation of this paper, the proof process is not presented, but it is actually similar to the proof of Theorem 1(2).

Hypotheses 1-2, Definitions 1-5, and Theorems 1-2 show that since interdomain path communication could involve the cooperative operation of different controllers and increasing the number of interdomain communications could increase the cost of cooperative interaction, the interdomain path cost is much greater than the intradomain path cost; the optional paths in each domain have sufficient redundancy to ensure connectivity between the network nodes within the domain, and for the multicast path across any two domains, only one interdomain link exists, reducing the multicast cost, such as that from network traffic. The subsequent analyses in this paper are based on this conclusion.

Finally, the specific form of the cost function of the optimal cross-domain multicast tree defined in this paper is obtained.

Assuming that the interdomain multicast tree and each intradomain multicast tree are the minimum cost trees, the minimum cost tree is defined in the same way as in Definition 5; i.e., each minimum cost tree can be defined as a set of paths $p_i (i = 1, ..., n)$ from the source node $src$ to $n$ destination nodes $d_i (i = 1, ..., n)$ and can be expressed as $T = \{p_1, \cdots, p_k, \cdots, p_n\}$. After the minimum cost $c_k$ of each path $p_k$ in $T$ is determined, the minimum cost of $T$ is obtained. If the construction cost of the intradomain multicast tree $T_i$ in each domain is expressed as $C_i$ and the number of destination nodes (if the domain has a BN for leaving the domain and the BN is also the destination node of the intradomain multicast tree $T_i$) within the domain $N_i$ is denoted by $n_i$, then we have:

$$C_i = \sum_{k=1}^{n_i} c_k \quad (10)$$

The calculation of the interdomain multicast tree cost $C_{int}$ is the same as that for an intradomain multicast tree. The total cost of the cross-domain multicast tree $cost(T_{cd})$ is the cost of the intradomain and interdomain multicast trees, and the total cost can be minimized as follows:

$$\min Cost(T_{cd}) = \sum_{i=1}^{m} C_i + C_{int} \quad (11)$$

where $\sum_{i=1}^{m} C_i$ represents the sum of the construction costs of all intradomain multicast trees, $m$ represents the number of domains, and $C_{int}$ represents the construction cost of the interdomain multicast tree.

## IV. SDWN WITH A MULTICONTROLLER INTELLIGENT CROSS-DOMAIN MULTICAST ROUTING ARCHITECTURE

The SDWN multiagent interdomain multicast routing strategy decomposes the construction of interdomain multicast trees into interdomain multicast trees and multiple intradomain multicast trees. Cross-domain multicast routing is achieved through multiagent collaboration. The architecture is illustrated in Fig. 7, and the detailed explanation is as follows.

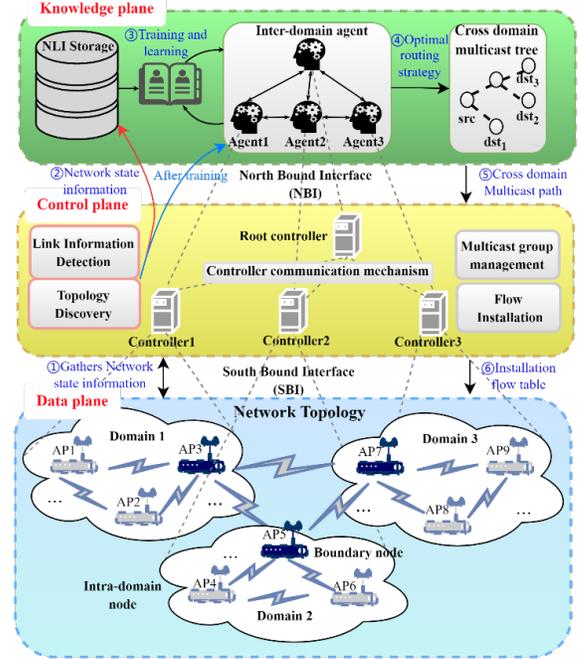

**Fig. 7.** SDWN with a multiagent cross-domain multicast routing architecture

① Each domain is equipped with a local controller in the control plane that periodically retrieves network state information for the corresponding domain. This information is synchronized to the root controller through a controller communication mechanism, and the root controller manages global network resources. ② The global network link information (NLI) collected in the controller plane is processed and stored in the knowledge plane. ③ Multiple agents in the knowledge plane learn and are trained on the basis of the NLI data from different time intervals through perception. ④ Interdomain agents select interdomain multicast trees, which involves selecting paths and boundary nodes among multiple domains. These decisions are then communicated to the agents in each controller domain to construct intradomain multicast trees. The construction of an optimal interdomain multicast tree is achieved through the collaboration of multiple agents. ⑤ The established interdomain multicast tree is synchronized with both the root and local controllers. Before network traffic arrives, flow tables are sent and installed by the control plane at the wireless access nodes in the data plane. The forwarding of traffic then occurs in the data plane.

### A. Data plane

The data plane is responsible for processing and forwarding network packets within the network, which consists of multiple network subdomains. Each domain comprises wireless access points (APs) and stations (STAs). APs form a multihop wireless network within each domain via wireless mesh technology [35]. APs are divided into intradomain nodes (INs) and boundary nodes (BNs), with each AP connecting to an STA within the domain. APs receive wireless data packets from terminal devices and, on the basis of instructions and policies provided by the control plane, select the optimal routing path to forward the packets from the source device to the destination device. Each domain periodically interacts with a corresponding local controller to transmit the wireless network link information from the current domain to the control plane.

*B. Control plane*

The control plane is responsible for collecting and analyzing network state information, formulating decisions, and distributing commands and policies to the APs in the data plane. It consists of a root controller and multiple local controllers. The local controllers communicate with the APs in the data plane through a southbound interface, synchronizing the network state information from their corresponding domains to the root controller via the controller communication mechanism (CCM). The root controller provides a global view of the network and manages and schedules global network resources. The control plane interacts with the knowledge plane through a northbound interface, facilitating the deployment and dissemination of policies in the knowledge plane. This approach includes network topology discovery, link information detection, a controller communication mechanism, multicast group management, and flow table installation.

Network topology discovery and link information detection are achieved by the controller sending data packets to the data plane to obtain relevant information. Network topology discovery involves periodically sending link layer discovery protocol (LLDP) [36] request packets, to which APs respond by encapsulating device port connections, IDs, and other information in reply packets. The controller then interprets the previous process and constructs the network topology on the basis of these packets. Link information detection is performed by sending PortStatsRequest packets to obtain port information. The controller decodes the reply message to obtain the following network link information: transmitted packets $tx_p$, received packets $rx_p$, transmitted bytes $tx_b$, received bytes $rx_b$, transmitted erroneous packets $tx_{err}$, received erroneous packets $rx_{err}$, and the duration of port transmission in terms of time $t_{dur}$. Using these parameters, the remaining bandwidth $bw_{ij}$, bandwidth $ubw_{ij}$, packet loss rate $loss_{ij}$, and error rate $err_{ij}$ of the network link can be calculated. The formulas for these calculations are as follows:

$$ubw_{ij} = \frac{\left|(tx_{bi} + rx_{bi}) - (tx_{bj} + rx_{bj})\right|}{t_{durj} - t_{duri}} \quad (12)$$

$$bw_{ij} = bw_{max} - ubw_{ij}$$

$$loss_{ij} = \frac{tx_{pi} - rx_{pj}}{tx_{pi}} \quad (13)$$

$$err_{ij} = \frac{tx_{erri} + rx_{errj}}{tx_{pi} + rx_{pj}} \cdot 100\% \quad (14)$$

where $tx_{b*}$, $rx_{b*}$ and $t_{dur*}$ represent the transmitted byte count, received byte count, and duration of transmission for nodes, respectively. Similarly, $tx_{p*}$, $tx_{err*}$, $rx_{p*}$, and $rx_{err*}$ represent the transmitted packet count, erroneous packet count, received packet count, and erroneous packet count, respectively, for node *.

In an SDN [37], since communication between two switches needs to be forwarded through a controller, the network link delay between two switches needs to be approximately calculated. The controller calculates the round-trip time (RTT) from the controller to the two switches, denoted as $RTT_1$ and $RTT_2$, by analyzing the timestamp information for the packets passing through the link. Additionally, on the basis of the LLDP messages, the controller calculates the forward propagation time $T_{fwd}$ and reply propagation time $T_{re}$ between the switches. The link delay $delay_{ij}$ is approximately calculated using these values, as shown in (15).

$$delay_{ij} = \frac{(T_{fwd} + T_{re} - RTT_1 - RTT_2)}{2} \quad (15)$$

Furthermore, the distance $dist_{ij}$ between two APs is calculated on the basis of the deployment coordinates of the wireless APs. The network link information (NLI) obtained from the abovementioned calculations is stored in the NLI storage of the knowledge plane after max-min normalization [34].

The controller communication mechanism (CCM) ensures stable and fast communication between the local and root controllers. The MBGP is used in traditional SDN multidomain communication. However, configuring this protocol in an SDN environment can be complex and may lead to inconsistent messages among multiple controllers during message updates. Therefore, a RESTful API design is proposed to address these issues, and a communication mechanism between the local controllers and the root controller is established. The CCM offers flexibility, allowing the root and local controllers to interact in a unified manner and adapt to network environments and requirements. The RESTful API is a standardized approach with well-defined specifications and constraints, reducing compatibility issues. Moreover, it provides scalability, enabling easy expansion and enhancing communication capabilities. The CCM also achieves loose coupling, allowing each controller to be developed and deployed independently, which improves maintainability and scalability. Furthermore, this mechanism is portable and can run on different environments and platforms, enhancing its versatility and adaptability.

The main functionality of multicast group management (MGM) is assessing domains in which the multicast source and multicast destination nodes are located and managing the joining and leaving of multicast group members. The MGM analyzes network traffic and packet information to determine the domains with multicast sources and destinations; it coordinates communication among different domains to ensure the proper transmission of multicast flows across domains and communicates with controllers in each domain, providing them with relevant information about multicast flows and coordinating the transmission paths of multicast flows across domains. Additionally, the MGM can handle join and leave requests from multicast users and update the multicast tree accordingly.

(1) Node joining: When a node $v_i$ in domain $N_k$ sends a dynamic join request $req_k = (v_i, add)$, the MGM retrieves the current multicast tree $T_k$ and network topology information for domain $N_k$. If $v_i$ is not already in $T_k$, a request is sent to the agent in the current domain to construct the minimum-cost path $p_i$ from $v_i$ to $T_k$. $T_k$ is then updated as $T_k \leftarrow T_k \cup p_i$. Simultaneously, $v_i$ is marked as an online multicast group member capable of receiving data.

(2) Node leaving: When a multicast group member $v_i$ in $N_k$ sends a node leave request $req_k = (v_i, leave)$, the MGM retrieves the current multicast tree $T_k$. It identifies the path $p_i$ from $v_i$ to $T_k$ and performs pruning operations. $T_k$ is updated as $T_k \leftarrow T_k - p_i$, the corresponding flow tables are removed, and $v_i$ is marked as an offline node for which data transmission ceases.

The flow table receives instructions from the knowledge plane through the northbound interface of the SDN and installs multicast flow entries. When receiving node join and leave requests, it modifies or deletes existing flow table entries to facilitate the addition or removal of group members, ensuring accurate data forwarding.

*C. Knowledge plane*

The knowledge plane, an essential addition to the SDN architecture, serves as a crucial component in the multiagent deep reinforcement learning intelligent intradomain multicast routing algorithm proposed in this paper. It includes NLI storage, which stores network link information collected by the control plane from the data plane. The NLI must be processed into a traffic matrix (TM) to provide inputs for the training of intelligent agents and subsequent learning. These agents are trained to collaborate in the construction of interdomain multicast trees. Once trained, they issue instructions to the control plane, which then sends the corresponding flow table entries to the data plane.

On the basis of the description and modeling of the interdomain multicast problem in Section III of this paper, we obtain a solution by constructing a minimum-cost Steiner tree. We decompose the interdomain multicast tree construction problem into multiple subproblems, encompassing interdomain multicast tree and multiple intradomain multicast tree construction. To differentiate from the intelligent agents responsible for constructing intradomain multicast trees, we refer to the intelligent agents involved in interdomain multicast tree construction as interdomain agents. The state space and reward function of interdomain agents are the same as those of intradomain agents, but their action spaces differ. The action space of interdomain agents consists of a set of connecting edges between different domains. In contrast, the action space of intradomain agents consists of a set of nodes within a domain. This distinction arises because the domains are abstracted as nodes when interdomain multicast trees are constructed. However, multiple connecting edges could exist between adjacent domains, making the use of a node set as the action space inappropriate. The design is described in detail in Section V of this paper. Now, we illustrate the process of multiple intelligent agents collaborating to construct an interdomain multicast tree via two simple examples.

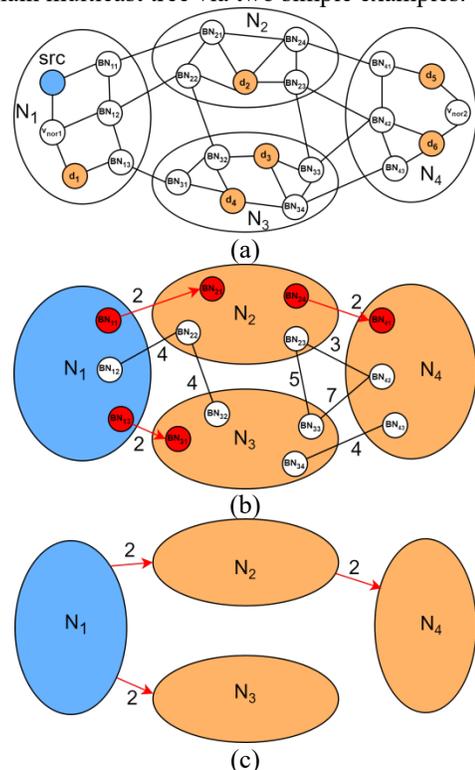

Figure 8 Construction of an interdomain multicast tree

Example 1: Fig. 8 depicts a multidomain network topology with four domains. The adjacent domains are connected through BNs, and an interdomain multicast tree is established. In Fig. 8(a), the source and destination nodes are labeled and located in different domains. In Fig. 8(b), we abstract the domains as nodes, where $N_1$ represents the source domain and where $N_2, N_3,$ and $N_4$ represent the target domains. We retain only the edges connecting adjacent domains and the boundary nodes, assigning a weight to each edge. The goal is to minimize the overall cost. The interdomain agents select four edges, $(BN_{11}, BN_{21})$、$(BN_{13}, BN_{21})$、$(BN_{24}, BN_{41})$, and $(BN_{34}, BN_{43})$, to connect the source domain and the target domains. This construction of the interdomain multicast tree results in the minimum cost, $C_{int} = 6$, and the boundary nodes required for each domain to reach other domains are determined. The final constructed interdomain multicast tree, $T_{int}$, is depicted in Fig. 8(c).

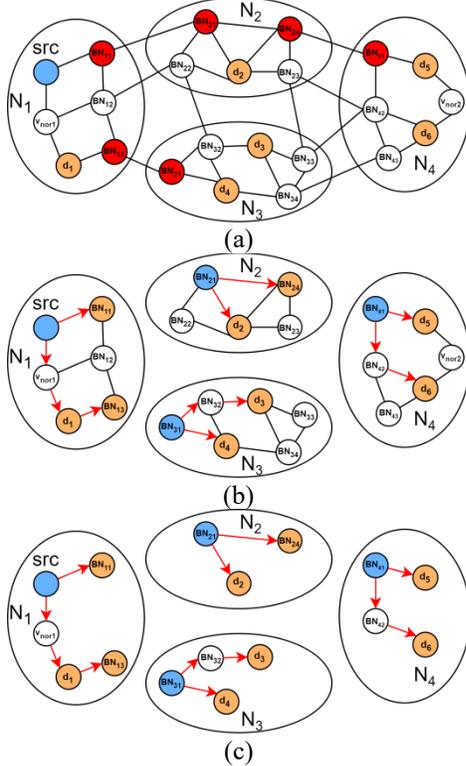

Figure 9 Construction of multicast trees within the domain

Example 2: Fig. 9 illustrates the process of constructing intradomain multicast trees. We begin by constructing the interdomain multicast tree and selecting the boundary nodes. In Fig. 9(a), on the basis of the interdomain multicast tree constructed in Fig. 8, we highlight the selected boundary nodes in red. These boundary nodes can be considered source and destination nodes within each domain. In Figure 9(b), we designate $BN_{11}$ and $BN_{13}$ as the destination nodes for domain $N_1$; $BN_{21}$ as the source node; $BN_{24}$ as the destination node for domain $N_2$; $BN_{31}$ as the source node for domain $N_3$; and $BN_{41}$ as the source node for domain $N_4$. Then, the corresponding intradomain agents construct the intradomain multicast trees, $T_1, T_2, T_3$, and $T_4$, from the respective source nodes to destination nodes within each domain. The final constructed intradomain multicast trees for each domain are depicted in Fig. 9(c).

The two examples above demonstrate the construction process of interdomain multicast trees and multiple intradomain multicast trees. Initially, the interdomain agents construct the interdomain multicast tree. Once the boundary nodes for each domain are determined, the intradomain multicast tree is constructed for each domain. Finally, by combining the interdomain multicast tree with the multiple intradomain multicast trees, we obtain the interdomain multicast tree, as depicted in Fig. 10.

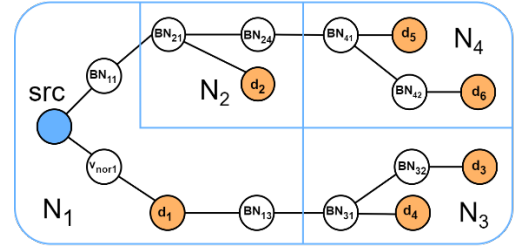

Figure 10 Cross-domain multicast tree

## V. MA-CDMR Algorithm

On the basis of the design of the multicast tree-based solution approach in Section 4.3 for the previous knowledge plane, this paper presents a process diagram of the MA-CDMR algorithm used to obtain interdomain multicast tree-based solutions, as shown in Fig. 11. First, the global network topology and link information are obtained through the multicontroller SDWN architecture. Then, this network information is stored in the NLI storage of the knowledge plane, providing data for intelligent agent training and learning. Second, the interdomain agents perceive the global network information and determine the optimal interdomain multicast tree, thereby identifying the boundary nodes for each domain to neighboring domains. Next, the information is synchronized to the intradomain agents of each domain, and the intradomain agents of the respective domains construct the optimal intradomain multicast trees. Finally, the optimal interdomain and multiple intradomain multicast trees are combined to generate the best cross-domain multicast tree, obtaining the forwarding paths for optimal interdomain routing in the global network.

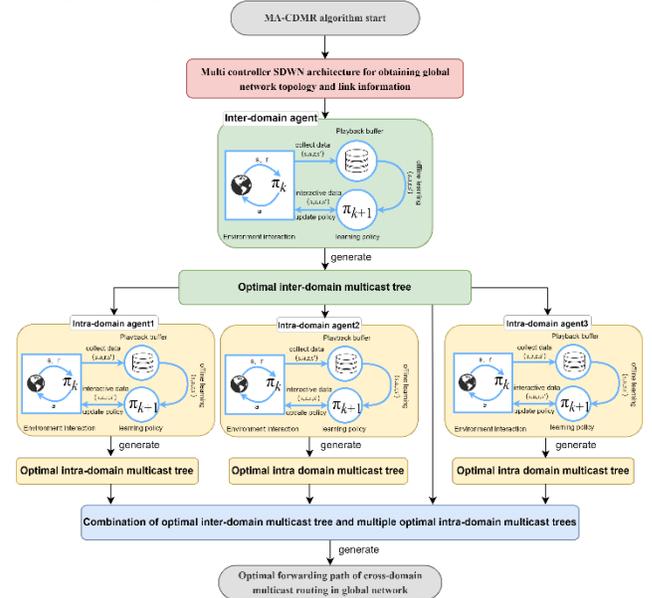

Figure 11 MA-CDMR algorithm flowchart

The MA-CDMR algorithm is described in detail below, including the design of the agent state space, action space, reward function, and training strategy of multiple agents. Finally, the pseudocode of the algorithm is introduced.

*A. Design of agents*

In the MA-CDMR algorithm, we design two agents: interdomain agents for constructing interdomain multicast trees and intradomain agents for constructing intradomain multicast trees. There is only one interdomain agent, while each domain has an intradomain agent. The state space and reward function are the same for both types of agents, but the action space differs. A detailed explanation is provided below.

1) State space

The state space of an agent is related to its ability to perceive and understand the environment. It should include sufficient environmental information to enable the agent to make meaningful decisions. Therefore, we construct a multichannel matrix $X = [bw, delay, loss, err, dist]$ to reflect the agent's state $s$, consisting of the remaining bandwidth $bw$, $delay$, packet loss rate $loss$, error rate $err$, and distance $dist$ of the network links, as well as the state matrix $M_T$ representing the construction status of the multicast tree. The multichannel state matrix is illustrated in Fig. 12.

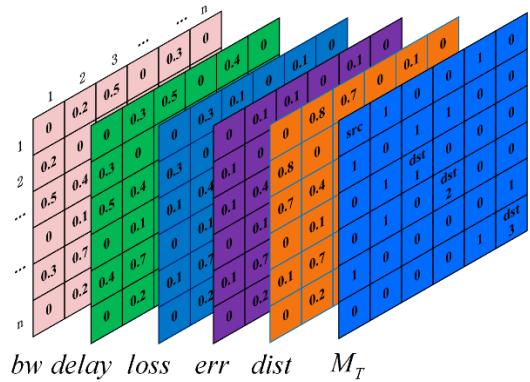

Figure 12 State matrix of an agent

The set of all possible variations of the multichannel matrix $X$ forms the state space $S$ of the agent, which can be represented as $S = \{X_1, X_2, ..., X_n\} = \{s_1, s_2 .... s_n\}$. For example, if a particular edge is selected to be included in the multicast tree, the agent's state changes from $s_t$ to $s_{t+1}$. By designing the state space in this way, meaningful information about network links and nodes and the specific progress of multicast tree construction is provided. This enables the agent to make decisions on the basis of the current network conditions and select the optimal path for constructing the multicast tree.

2) Action space

The action space determines the range and granularity of an intelligent agent's operations. The design of the action space should encompass all possible actions that the agent needs to take and be adaptable to the environmental characteristics and the task requirements. Therefore, we design action spaces separately for interdomain agents and intradomain agents.

Action space for interdomain agents, denoted as $\mathcal{A}_{int}$: When constructing interdomain multicast trees, we abstract each domain as a node, and there may be multiple connecting edges between these nodes. Thus, the set of connecting edges between domains serves as the action space for interdomain agents, $\mathcal{A}_{int} = \{(BN_{11}, BN_{21}), (BN_{12}, BN_{32})\} \{(BN_{ia}, BN_{jb})\} = \{e_1, e_2, ..., e_k\}$, where $(BN_{ia}, BN_{jb})$ represents an edge connecting domain $N_i$ and domain $N_j$ and $BN_{ia}$ represents the a-th boundary node in domain $N_i$. Here, $k$ denotes the total number of edges between different domains. Only the edges connecting the current domain with neighboring domains are considered valid actions, whereas the remaining edges are invalid. This design enables the intelligent agent to select the most suitable boundary nodes and paths for interdomain multicast tree construction.

Action space for intradomain agents, denoted as $\mathcal{A}_{intra}$: The connectivity among intradomain nodes is complex, with numerous connecting edges. To reduce the complexity of the action space, we consider the set of intradomain nodes as the action space for intradomain agents, $\mathcal{A}_{intra} = \{v_1, v_2, ..., v_n\} = \{a_1, a_2, ..., a_n\}$, where each node $v$ corresponds to an action $a$. The valid actions are determined by the next-hop nodes from the current node, whereas the remaining actions are considered invalid actions. This design reduces ineffective exploration by intelligent agents and enables faster decision-making for determining the optimal path.

3) Reward function

Intelligent agents use the reward function to evaluate the positive or negative consequences of their actions during the learning process. It provides explicit feedback to guide an agent toward the desired goals. In our case, the optimization objectives are to maximize the remaining bandwidth of the multicast tree while minimizing the delay, packet loss rate, packet error rate, and distance between wireless access points (APs). We design the reward function on the basis of network link information parameters to achieve these goals. Since we design the action space on the basis of sets of links or nodes for a given state at different time steps, valid and invalid actions can be taken. When the agent performs a valid action $a$, transitioning from state $s_t$ to $s_{t+1}$, the process may generate two types of positive feedback and one type of negative feedback; executing an invalid action results only in negative feedback. Below, we design the reward function for different scenarios.

Suppose that adding the next-hop node or link after executing the current action $a_t$ only adds a regular node to the multicast tree. In this case, we design a single-step reward function $R_{part}$ to calculate the reward on the basis of the remaining bandwidth $bw_{ij}$, delay $delay_{ij}$, packet loss rate $loss_{ij}$, packet error rate $err_{ij}$, and distance $dist_{ij}$ between the added link $e_{ij}$ and the APs. The reward is calculated as shown in (16).

$$R_{part} = \beta_1 bw_{ij} + \beta_2 (1 - delay_{ij}) + \beta_3 (1 - loss_{ij}) \\ + \beta_4 (1 - err_{ij}) + \beta_5 (1 - dist_{ij}) \quad (16)$$

If adding the next-hop node or link leads to adding a destination node $d_k$ to the multicast tree, we design an end-task reward $R_{end}$ for this positive feedback. The reward is calculated on the basis of the remaining bandwidth $bw_k$, delay $delay_k$, packet loss rate $loss_k$, packet error rate $err_k$, and average distance $dist_k$ along the entire path from the source node $src$ to the destination node $d_k$. The reward is calculated as shown in (17).

$$R_{end} = \beta_1 bw_k + \beta_2(1-delay_k) + \beta_3(1-loss_k) \\ +\beta_4(1-err_k) + \beta_5(1-dist_k) \quad (17)$$

If the addition of the next-hop node or link leads to the formation of a loop in the multicast tree, we assign a penalty value $R_{loop} = C_1$ for this negative feedback.

Suppose that $a_t$ is an invalid action, meaning that it is neither the next-hop node for the current node nor a connecting edge between the current domain and neighboring domains. In this case, we also assign a penalty value $R_{hell} = C_2$ for this negative feedback.

4) Agent policy updating

In this algorithm, all intelligent agents utilize the actor–critic framework [38]. The framework consists of two components: the actor, represented by the policy network $\pi_\theta(a|s)$ with parameter $\theta$, and the critic, represented by the value network $V_\omega$ with parameter $\omega$. The actor interacts with the environment and learns a better policy via policy gradient updates guided by the critic's value function. The policy gradient update is performed as shown in (18).

$$\nabla_\theta J(\theta) = \mathbb{E}_{\pi_\theta}\left[\sum_{t=0}^{T} \psi_t \nabla_\theta \log \pi_\theta(a_t|s_t)\right] \quad (18)$$

where $T$ is the maximum number of steps for interacting with the environment, $\pi_\theta(a|s)$ is the policy function, $a_t$ denotes the current action, and $s_t$ corresponds to the current state. $\psi_t$ represents the temporal difference residual [39], which guides policy gradient learning; this variable is calculated via (19).

$$\psi_t = r_t + \gamma V_\omega(s_{t+1}) - V_\omega(s_t) \quad (19)$$

where $r_t$ represents the reward obtained at the current step, $\gamma$ is the discount factor, with a value between 0 and 1 ($\gamma \in [0,1]$), and $V_\omega(s_t)$ represents the expected value generated by executing action $a_t$ in state $s_t$

In the critic value network $V_\omega$, a learning approach that is based on the temporal difference residual is used. For an individual data point, the loss function in the value function is defined as shown in (20).

$$\mathcal{L}(\omega) = \frac{1}{2}\left(r + \gamma V_\omega(s_{t+1}) - V_\omega(s_t)\right)^2 \quad (20)$$

We can update the value function by using $r + \gamma V_\omega(s_{t+1})$ as the temporal difference target. However, this target does not provide the gradient to update the value function directly. Therefore, the gradient of the value function is calculated as shown in (21).

$$\nabla \mathcal{L}(\omega) = -\left(r + \gamma V_\omega(s_{t+1}) - V_\omega(s_t)\right)\nabla_\omega V_\omega(s_t) \quad (21)$$

B. Design of the multiagent training strategy

Training multiple intelligent agents is more complex than training a single agent because each agent interacts with the environment and directly or indirectly interacts with other agents. Furthermore, each agent continuously learns and updates its policy. Consequently, for each agent, the environment is nonstationary. On the other hand, in a multidomain environment, the intelligent agent in each domain has different training objectives, aiming to maximize rewards. This may require large-scale distributed training to improve efficiency. To address these challenges, this paper adopts a fully decentralized training approach known as independent learning (IL) [40].

In the MA-CDMR algorithm, for each agent within a domain, the global network environment is nonstationary, but the corresponding subdomain network environment is stationary. Policy execution for interdomain agents takes precedence over that for intradomain agents, and they operate in the context of the global network, resulting in a stationary environment. Therefore, by using IL training, each agent can independently learn in its environment without considering the changes made by other agents. Additionally, this approach exhibits good scalability as the network expands and the number of agents increases.

To increase the convergence speed of multiple intelligent agents and reduce training costs in large-scale network environments, the advantages of online reinforcement learning and offline reinforcement learning are combined, and a training method that integrates both is designed. To illustrate the differences between the training strategies proposed in this paper and three traditional training strategies, we present them in the form of a figure, as shown in Fig. 13.

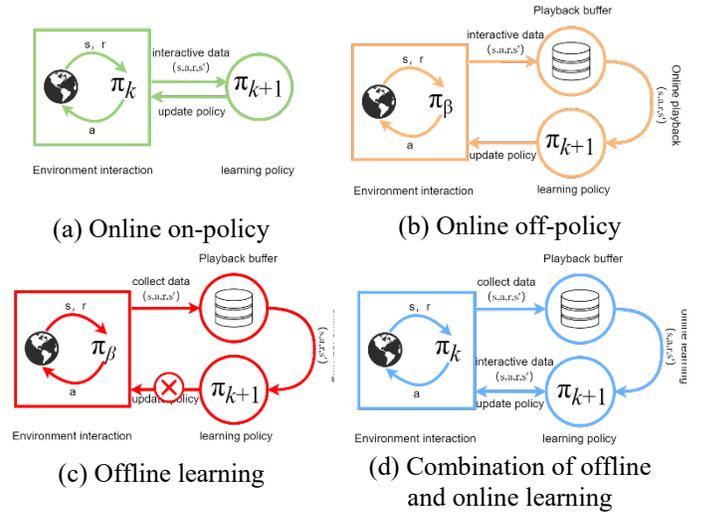

(a) Online on-policy  (b) Online off-policy

(c) Offline learning  (d) Combination of offline and online learning

Figure 13 Four training strategies

Figs. 13(a) and 13(b) depict on-policy and off-policy online reinforcement learning [41], respectively, both of which require real-time interactions with the environment. In on-policy learning, the policy used for interacting with the environment and updating data is the same policy $\pi_k$. On the other hand, in off-policy learning, the policy used for interacting with the environment and generating data is $\pi_\beta$, whereas the policy used for updating the data is $\pi_k$. Fig. 13(c) represents the training strategy of offline reinforcement learning [16]. It involves collecting a dataset before training the agent and then using it for offline training. Like off-policy learning, it follows an off-policy approach, but environmental interactions are not considered when updating the policy. Fig. 13(d) illustrates the training strategy designed in this paper, which combines the advantages of online and offline learning. It involves collecting a certain amount of data and using these offline data for training to establish an initial policy. Then, on-policy online reinforcement learning with environmental interactions is used to improve the policy in the actual environment. The hybrid training strategy of offline and online learning reduces the demand for online learning, improves data utilization efficiency, and can better handle dynamic changes in the actual environment.

*C. MA-CDMR algorithm design*

The MA-CDMR algorithm divides cross-domain multicast routing into two stages: interdomain multicast tree construction and intradomain multicast tree construction. In the first stage, interdomain agents perceive the global network environment and establish the interdomain multicast tree. This involves the selection of optimal interdomain forwarding paths and boundary nodes in each domain. The information is then synchronized with each intradomain agent to construct the intradomain multicast tree. Finally, the interdomain multicast tree and multiple intradomain multicast trees are combined to generate the best cross-domain multicast tree. For multicast user join and leave events, multicast path construction and tree pruning are performed accordingly. For a detailed implementation of the MA-CDMR algorithm, please refer to Algorithm 1.

The algorithm inputs are the multiple-domain global network topology $G(V,E)$, network link information NLI, a multicast node set $M = \{src\} \cup D$, and various hyperparameters for deep reinforcement learning. The algorithm outputs the optimal interdomain multicast tree from the source node to all destination nodes in each domain. Lines 1-3 denote the initialization of each agent's actor-network and critical network parameters and give the data buffers for intradomain and interdomain communication. The process in Line 4 is used to analyze the information collected by the control module, and the multicast group management module is applied to determine the domains where the source node and destination nodes reside. Lines 5-14 represent the function for the offline training of the agents. The functions on Lines 7-9 are used to retrieve a batch of reinforcement learning data with a batch size of $k$ from the collected offline data. The states $s_t$ and $s_{t+1}$ in the data are input into the critic network to compute $V_\omega(s_t)$ and $V_\omega(s_{t+1})$. Lines 10-11 are used to update the actor network and critic network parameters on the basis of the computed results. Line 13 returns the offline-trained actor network and the critic network. Line 17 generates the interdomain multicast tree state matrix $M_T^{int}$ on the basis of the domains where $src$ and $D$ are located. Line 18 stacks $M_T^{int}$ and $M_{NLI}^{int}$ to obtain the interdomain agent state $s_t$, which enters the loop of the interdomain agents. Line 20 samples an action $a_t$ from the action set output by the interdomain agent on the basis of $s_t$. Executing action $a_t$ yields a reward value $r_t$ and the next state $s_{t+1}$. The function in Line 21 is used to collect offline data $(s_t, a_t, r_t, s_{t+1})$ and store them in the interdomain data buffer $B_{int}$. Line 22 indicates offline training for the interdomain agents without interacting with the environment. Lines 23-24 synchronize the constructed interdomain multicast tree to each intradomain agent; additionally, the forwarding boundary nodes are located for each domain, and the intradomain agent loop is entered. Lines 26-27, for each intradomain agent, are used to construct the intradomain multicast state $M_T^{intra}$ on the basis of $BN$, $src$, and $D$. Stacking $M_T^{intra}$ and $M_{NLI}^{intra}$ in each case yields the state $s_t$. In Line 28, each intradomain agent samples an action $a_t$ from the action set output on the basis of $s_t$. Executing action $a_t$ yields a reward value $r_t$ and the next state $s_{t+1}$. Lines 29-30 collect offline training data for intradomain agents, and the data are stored in the corresponding intradomain data buffer $B_{intra}$. Line 31 involves offline training for each intradomain agent without interacting with the environment. In Lines 32-34, each intradomain agent determines whether a corresponding intradomain multicast tree $T_{intra}$ is constructed. Once completed, a loop exits; otherwise, the corresponding intradomain agent updates the state $s_t \leftarrow s_{t+1}$. In Lines 35-36, intradomain agents interact with the environment, update the actor and critic network parameters online and improve the policy learned from offline training. In Lines 40-46, if the final interdomain multicast tree has not yet been established, the interdomain agents update the state $s_t \leftarrow s_{t+1}$ and interact with the environment, performing online training to adjust the interdomain strategy. Lines 47-50 are used to determine whether the interdomain multicast tree and all intradomain multicast trees are constructed. If not, the loop continues, and if completed, $T_{int}$ and $T_{intra}$ are combined to construct the cross-domain multicast tree; then, the loop proceeds to the next round, $ep$.

**Algorithm 1 MA-CDMR**

**Require:** global network topology $G(V, E)$, network link information $NLI$, Multicast node set $M = src \cup D$, weight factor $\beta_l, l = 1, 2, ...5$, actor network learning rate $\alpha_1$, critic network learning rate $\alpha_2$, decay factor $\gamma$, batch-size for offline training $k$, update frequency $N_{update}$, number of training episodes $episodes$, number of domains $m$

**Ensure:** The optimal cross-domain multicast tree from the source node to all destination nodes in each domain. $tree(src, D)$

1: Initialize inter-domain and intra-domain Actor network $\theta_{int}, \theta_{intra}$
2: Initialize inter-domain and intra-domain Critic network $\omega_{int}, \omega_{intra}$
3: Initialize inter-domain and intra-domain data buffers $B_{int}, B_{intra}$
4: Analyze the source node and destination node domains based on the multicast group management module
5: **function** OFFLINE_TRAINING($B$)
6:    **for** $batch$ in $B$ **do**
7:       Sample a batch of data $(s_t, a_t, r_t, s_{t+1})$ with a batch size of $k$ and calculate the policy $\pi_\theta(a|s)$ and value function $V_\omega(s)$ for each state
8:       Select action $a$ based on the policy $\pi_\theta(a|s)$ and obtain $r$ and $s_{t+1}$ from the offline data
9:       Calculate $V_\omega(s_t)$ and $V_\omega(s_{t+1})$ by inputting $s_t$ and $s_{t+1}$ into the Critic network. Then, calculate the temporal difference residual based on $r$, $V_\omega(s_t)$, and $V_\omega(s_{t+1})$
10:      Update the Critic network parameters $\omega$ to minimize the mean squared error loss: $\omega \leftarrow \omega - \alpha_2 \cdot \nabla \mathcal{L}(\omega)$
11:      Update the Actor network parameters $\theta$ to maximize the objective function: $\theta \leftarrow \theta + \alpha_1 \cdot \nabla_\theta J(\theta)$
12:    **end for**
13:    Return the trained Actor network and Critic network
14: **end function**
15: **for** $ep \leftarrow 1$ to $episodes$ **do**
16:    **for** Network link information matrix $M_{NLI}^{int}, M_{NLI}^{intra}$ in NLI Storage **do**
17:       Reset the environment based on the domains where $src$ and $D$ are located to obtain inter-domain environment $M_T^{int}$
18:       Stack inter-domain environment $M_T^{int}$ and network link information $M_{NLI}^{int}$ to obtain the state $s_t$ of the inter-domain agent
19:       **while** True **do**
20:          The inter-domain agent samples an action $a_t$ from the set of actions outputted by $s_t$. Execute $a_t$ to obtain $r_t$ and $s_{t+1}$
21:          The inter-domain agent stores $(s_t, a_t, r_t, s_{t+1})$ in $B_{int}$
22:          Collect data for a certain period of time into $B_{int}$ and perform offline training using $offline\_training(B_{int})$
23:          **if** $T_{int}$ has been constructed **then**
24:             The information is synchronized to the intra-domain agents to determine the forwarding boundary nodes $BN$ for each domain
25:             **while** True **do**
26:                Each intra-domain agent constructs its own intra-domain $M_T^{intra}$ based on $BN$, $src$, and $D$
27:                Each intra-domain agent stacks its own $M_T^{intra}$ and $M_{NLI}^{intra}$ to obtain the $s_t$
28:                Intra-domain agents sample an $a_t$ from the set of actions outputted by their respective $s_t$. Execute $a_t$ to obtain $r_t$ and $s_{t+1}$
29:                Store $(s_t, a_t, r_t, s_{t+1})$ in the corresponding intra-domain data buffer $B_{intra}$
30:                Each intra-domain agent collects data for a certain period of time into $B_{intra}$
31:                Perform offline training using offline_training($B_{intra}$) for each intra-domain agent.
32:                Each intra-domain agent constructs an $T_{intra}$ specific to its domain.
33:                **if** $T_{intra}$ has been constructed **then break**
34:                The update of the intra-domain agent state $s_t \leftarrow s_{t+1}$
35:                **for** $i \leftarrow 1$ to $n_{update}$ **do**
36:                   Online update $\omega_{intra} \leftarrow \omega_{intra} - \alpha_2 \cdot \nabla \mathcal{L}(\omega_{intra})$
37:                   Online update $\theta_{intra} \leftarrow \theta_{intra} + \alpha_1 \cdot \nabla_\theta J(\theta_{intra})$
38:                **end for**
39:             **end while**
40:          **else** //The $T_{int}$ has not been fully constructed yet
41:             The update of the inter-domain agent state $s_t \leftarrow s_{t+1}$
42:             **for** $i \leftarrow 1$ to $n_{update}$ **do**
43:                Online update $\omega_{int} \leftarrow \omega_{int} - \alpha_2 \cdot \nabla \mathcal{L}(\omega_{int})$
44:                Online update $\theta_{int} \leftarrow \theta_{int} + \alpha_1 \cdot \nabla_\theta J(\theta_{int})$
45:             **end for**
46:          **end if**
47:          **if** $T_{int}$ and all $T_{intra}$ has been completed **then**
48:             The $T_{cd}$ is constructed by combining $T_{int}$ and multiple $T_{intra}$
49:             break
50:          **end if**
51:       **end while**
52:    **end for**
53: **end for**

## VI. EXPERIMENT AND ANALYSIS

This section focuses primarily on the experiments and analyses of the results. First, the experimental setup is introduced. Then, the hyperparameters used in the experimental process are discussed and set. Finally, comparative experiments are conducted with classic multicast tree optimization methods such as KMB and SCTF and algorithms such as DRL-M4MR and MADRL-MR that utilize reinforcement learning to construct multicast trees. The results of the comparative experiments are discussed individually.

### A. Experimental environment

The server software system used in the experiment is Ubuntu 18.04.6, with a hardware configuration consisting of a 64-core processor and a GeForce RTX 3090 graphics card. Mininet-WiFi 2.3.1b [42] is installed on the server to support the simulation platform for the SDWN. A Ryu 4.3.4 controller is used [43]. Real network traffic is simulated via the Iperf [44] tool.

The experimental setup employs a multidomain network topology, as shown in Fig. 14. The link parameters of the global network follow a uniform distribution and are randomly generated within specific ranges. The random ranges for the link bandwidth and delay are 5-40 Mbps and 1-10 ms, respectively. The distances between wireless access points (APs) are set within a 30-120 m range. The distribution of simulated traffic generation is illustrated in Fig. 15.

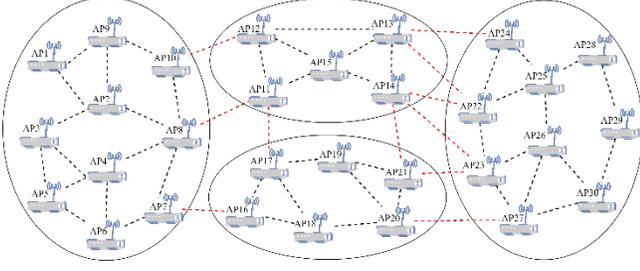

Figure 14 Multidomain network topology

*B. Performance metrics*

We evaluate our algorithm with thoroughness, considering the convergence of the agent's reward value and the following metrics of the multicast tree: bottleneck bandwidth, delay, packet loss rate, length, and average distance between wireless APs.

(1) A description of the reward value can be found in Section V-A, and all analyses explicitly refer to the equation provided in that section.

(2) To evaluate the remaining bandwidth, delay, packet loss rate, packet length, and average distance between wireless APs in the multicast tree, we perform multiple measurements and use the average values as evaluation metrics. The corresponding equations are shown in (22).

$$\overline{bw}_{tree} = average \frac{\sum_n \sum_{p_k \in tree} bw_k}{n \cdot K}$$

$$\overline{delay}_{tree} = average \frac{\sum_n \sum_{p_k \in tree} delay_k}{n \cdot K}$$

$$\overline{loss}_{tree} = average \frac{\sum_n \sum_{p_k \in tree} loss_k}{n \cdot K} \quad (22)$$

$$\overline{len}_{tree} = average \frac{\sum_n len_{tree}}{n}$$

$$\overline{dist}_{tree} = average \frac{\sum_n \sum_{p_k \in tree} dist_k}{n \cdot K}$$

where $\overline{bw}_{tree}$, $\overline{delay}_{tree}$, and $\overline{loss}_{tree}$ represent the average bottleneck bandwidth, delay, and packet loss rate, respectively, of the multicast tree obtained over multiple measurements. $\overline{len}_{tree}$ and $\overline{dist}_{tree}$ represent the average length and the distance between wireless Aps in the multicast tree, respectively, obtained from multiple measurements. $bw_k$, $delay_k$, $loss_k$, and $dist_k$ represent the bottleneck bandwidth, delay, packet loss rate, and distance, respectively, of path $p_k$ from the source node to different destination nodes within the multicast tree. $n$ represents the number of measurements taken at a specific moment, and $K$ represents the number of paths $p_k$.

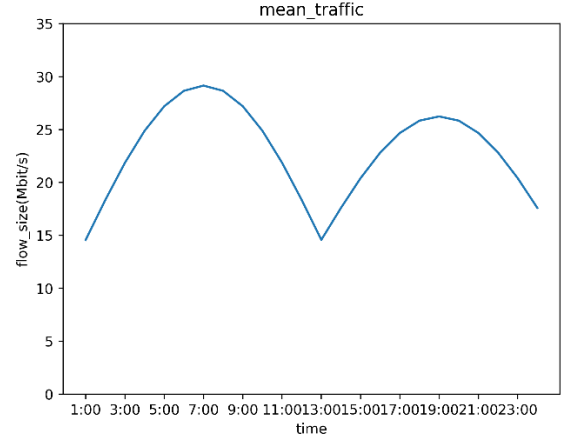

Figure 15 Simulated flow distribution

*C. Experimental parameter settings*

We utilize the actor–critic network as the core framework for our intelligent agents. Initially, each agent is trained via online learning with an on-policy approach. To validate the effectiveness of our designed offline–online learning combination training method, we conduct comparative experiments between agents trained with online on-policy learning and agents trained with the offline–online combined learning. The results of these experiments are depicted in Fig. 16.

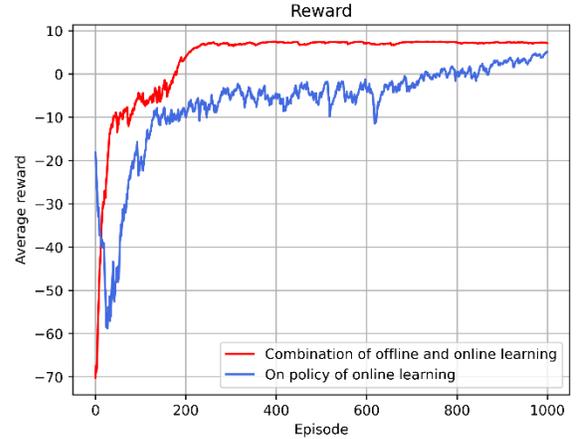

Figure 16 Agent training strategy settings

The results shown in Fig. 16 indicate that online reinforcement learning with an on-policy training approach tends to converge, but the convergence speed is relatively slow. This is because, in this training approach, the agents must interact with the environment in real time and continuously update their policies, thus requiring a long time to achieve convergence. On the other hand, using our designed offline–online combined learning approach results in a faster convergence speed, with the agents reaching convergence values in a shorter training time. This is achieved by reducing the interaction between the agents and the environment. We first perform offline training via a precollected dataset and then consider environmental interactions to adjust the agents' policies. This reduces the training cost associated with

agent–environment interactions and accelerates the convergence speed of the agents.

During the training process of the deep reinforcement learning model, the settings of the hyperparameters significantly impact the experimental results. In our experiment, we select a group of cross-domain multicast nodes with complex interdomain and intradomain route options as representatives, denoted as $M = \{src\} \cup D = \{3\} \cup \{6,15,18,19,26,28\}$. First, we set the positive feedback reward values $R_{part}$ and $R_{end}$ and the negative feedback penalty values $R_{hell}$ and $R_{loop}$. On the basis of our previous work focused on MADRL-MR [12], we conduct numerous experiments to determine the optimal weight factors for parameters such as the remaining bandwidth, delay, packet loss rate, incorrect packet rate, and distance between APs, which are found to be [0.7,0.3,0.1,0.1,0.1]. Additionally, we conduct experiments on the negative feedback penalty values $R_{hell}$ and $R_{loop}$ and find that $R_{hell} = -0.7$ and $R_{loop} = -0.5$ yield the best results. Building upon these findings, we modify the proportions of $R_{part}$ and $R_{end}$ to enable the agents to reach the destination nodes as quickly as possible. Finally, we conduct many experiments based on this modification, and the results are depicted in Fig. 17.

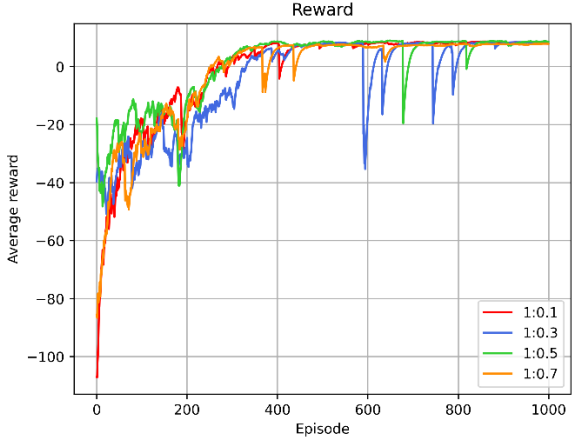

Figure 17 Reward comparison for four $R_{part}$ and $R_{end}$ ratios

In the experiments, we conducted trials with four different reward ratio settings: 1:0.1, 1:0.3, 1:0.5, and 1:0.7. We also explored a 1:1 reward ratio but found that the agent-related functions did not converge. Setting the reward ratio for single-step rewards at a reasonable rate is essential, as this will help the agents take as few steps as possible to reach the destination node. The results in Fig. 17 indicate that the 1:0.1 reward ratio produces the best performance.

Next, the learning rates ($\alpha_1$ and $\alpha_2$) of the actor and critic networks play crucial roles in the training process and network performance. $\alpha_1$ is related to the speed and magnitude of policy updates, whereas $\alpha_2$ is related to the speed and magnitude of value function updates. High learning rates can accelerate policy updates and the agent convergence speed but they also bring the risk of instability and overfitting. On the other hand, low learning rates enhance stability but at the cost of slower learning, requiring more training iterations to converge. Therefore, we conducted experiments to navigate these trade-offs and determine the optimal settings for $\alpha_1$ and $\alpha_2$, as depicted in Fig. 18.

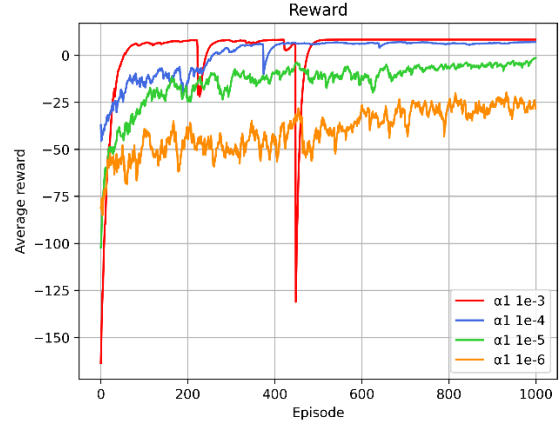

(a) learning rate $\alpha_1$

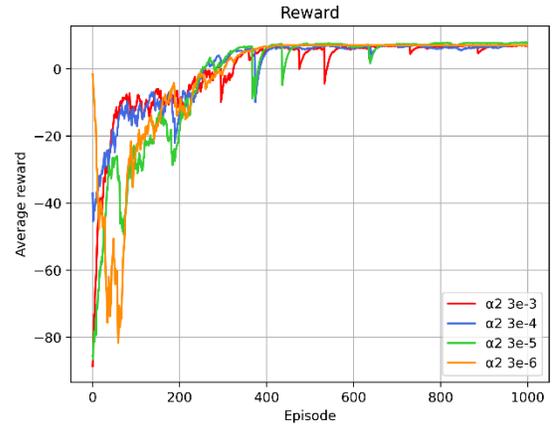

(b) learning rate $\alpha_2$

Figure 18 Learning rate settings

In the experiments conducted for $\alpha_1$, we initially set $\alpha_2$ to 3e-4. The results shown in Fig. 18(a) indicate that the best convergence is achieved when $\alpha_1$ is set to 1e-4. Following this, while keeping $\alpha_1$ at 1e-4, we adjusted the value of $\alpha_2$. Fig. 18(b) shows that the best convergence is achieved when $\alpha_2$ is set to 3e-4.

Next, we considered the setting of the reward discount factor $\gamma$. The discount factor relates to the relative importance of immediate and future rewards. A high discount factor emphasizes future rewards, encouraging the agent to consider long-term decision-making. However, this may result in slow convergence. Conversely, a low discount factor corresponds to the prioritization of immediate rewards, potentially leading to faster training but compromising long-term planning capabilities. In our experiments, we explored different values of $\gamma$ to determine their effects, and the results are shown in Fig. 19. The results indicate that when $\gamma$ is set to 0.9, the convergence speed is slow, but the reward values stably converge.

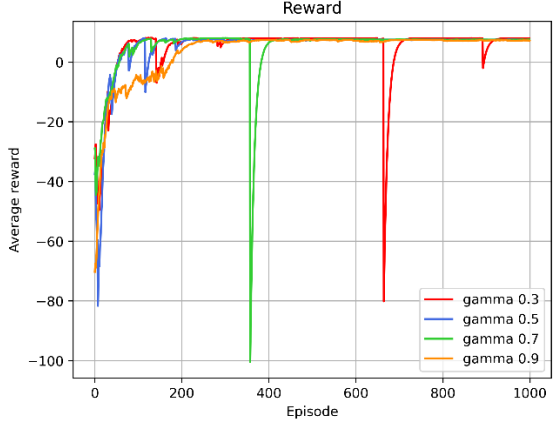

Figure 19 Attenuation factor settings

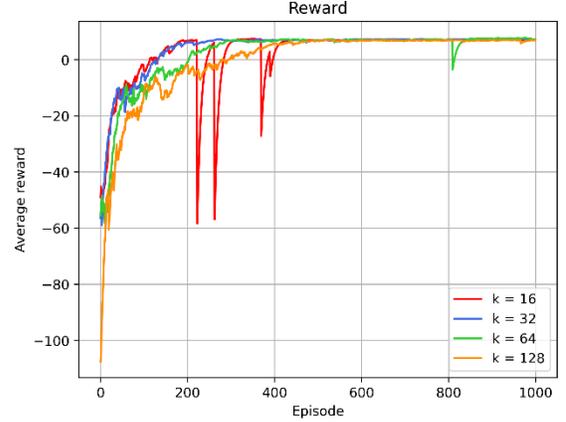

Figure 21 Batch size setting

Next, we consider the update frequency setting for online learning, denoted as $n_{update}$. A high value of $n_{update}$ can accelerate the convergence speed, improve training stability, reduce the number of required iterations, and decrease fluctuations. However, setting $n_{update}$ too high may increase the computational overhead and the risk of overfitting. In our experiments, we investigated different values of $n_{update}$ to determine their effects, and the results are shown in Fig. 20.

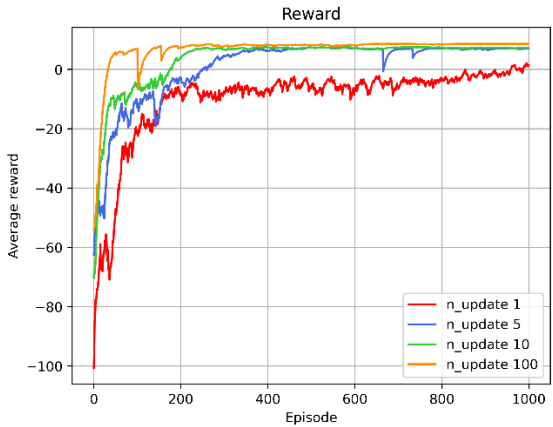

Figure 20 Update frequency

When $n_{update} = 1000$ is selected, the agents are unable to converge. Therefore, we adjusted the value of $n_{update}$ to 1, 5, 10, and 100. The best performance was achieved when $n_{update} = 100$. However, when $n_{update} = 10$, the convergence speed was slightly slower than when this parameter was set to 100. To prevent overfitting, we ultimately set $n_{update}$ to 10.

Finally, we considered batch size $k$ settings in offline learning. In offline reinforcement learning, the dataset is typically divided into batches for training, and the batch size $k$ is a vital hyperparameter. A larger value of $k$ can accelerate the training speed and reduce the variance of parameter updates, but it also increases memory consumption. Furthermore, a large $k$ may result in overfitting for training data, resulting in poor generalization performance. In our experiments, we explored different values of $k$ to determine their effects, and the results are shown in Fig. 21.

The results shown in Fig. 21 indicate that setting $k$ to 16, 32, 64, and 128 leads to similar convergence in terms of reward values. However, when $k$ is set to 32, the convergence speed is fastest, and the overall convergence performance is best. Therefore, we set $k$ to 32 on the basis of these findings.

### D. Comparative experiments and results

To evaluate the performance of the proposed MA-CDMR algorithm, we compared it with four other SDN multicast methods, namely, the classic multicast tree optimization methods KMB and SCTF, as well as the recently proposed reinforcement learning-based multicast routing methods DRL-M4MR and MADRL-MR. Here, we provide an expanded description of each of the comparative algorithms:

- KMB: KMB is an algorithm that Kou, Markousky, and Berman proposed in 1981 [6]. It is used to solve the Steiner tree problem and minimizes the distance between edges.
- SCTF: Selective closest terminal first (SCTF) is a method that can be directly applied to directed graphs. In [17], the concept of SCTF was used to solve a multicast tree problem.
- DRL-M4MR: DRL-M4MR is a method introduced in [11]. It uses the deep Q-network (DQN) reinforcement learning algorithm to compute multicast routing paths.
- MADRL-MR: MADRL-MR is a method presented in [12] and designed to address multicast routing in a single-domain environment. It decomposes the multicast task into multiple subtasks, which multiple agents collaboratively complete in the same environment to construct a multicast tree.

We implemented the SDN multicast algorithms mentioned above in a multidomain environment. Specifically, given the source and destination nodes, each algorithm generated multicast trees across domains. We calculate the average bottleneck bandwidth, average delay, average packet loss rate, number of links, and average link distance of the corresponding paths. The calculation methods for these performance metrics can be found in Section VI-B. These metrics are subsequently averaged over each time step to obtain the average bottleneck bandwidth, average delay, average packet loss rate, average multicast tree length, and average link distance.

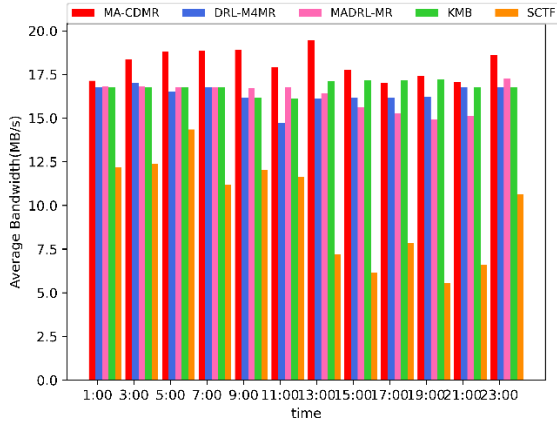

(a) Average bottleneck bandwidth

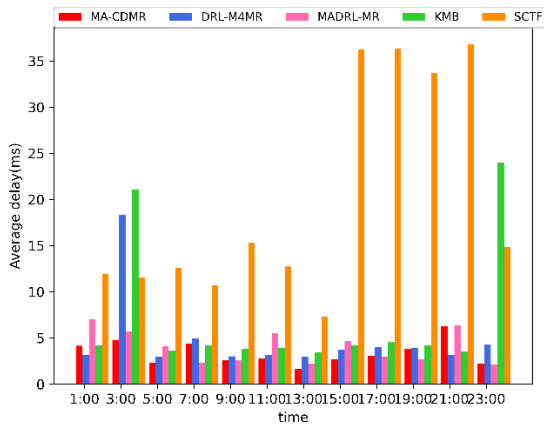

(b) Average delay

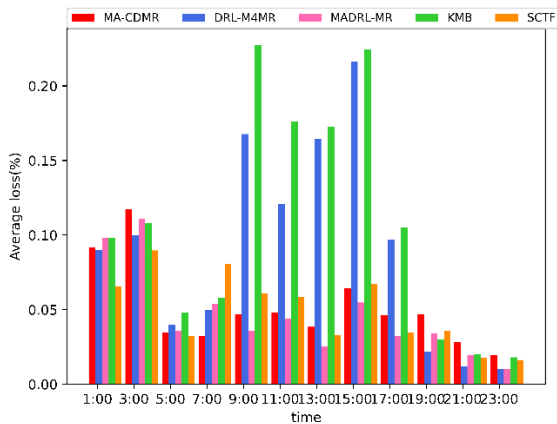

(c) Average packet loss rate

Figure 22 Comparison of the bandwidth, delay and, packet loss rate of the cross-domain multicast tree

Fig. 22(a) compares the average bottleneck bandwidths of the multicast trees constructed by using MA-CDMR with those of DRL-M4MR, MADRL-MR, KMB, and SCTF. The results indicate that MA-CDMR significantly outperforms SCTF in terms of the average bottleneck bandwidth, with an average improvement of 46.01%. It slightly outperforms DRL-M4MR, MADRL-MR, and KMB, with improvements of 9.61%, 10.11%, and 7.09%, respectively.

Fig. 22(b) compares the average delay of the multicast trees constructed via MA-CDMR with those of DRL-M4MR, MADRL-MR, KMB, and SCTF. The results show that MA-CDMR achieves a lower average delay than KMB and SCTF, with average improvements of 26.39% and 78.74%, respectively. Compared with DRL-M4MR and MADRL-MR, MA-CDMR achieves performance improvements of 12.47% and 7.17%, respectively; therefore, MA-CDMR exhibits favorable performance regarding average delay.

Fig. 22(c) compares the average packet loss rates of the multicast trees constructed via MA-CDMR with those of DRL-M4MR, MADRL-MR, KMB, and SCTF. The results indicate that all algorithms achieve low average packet loss rates. MA-CDMR outperforms DRL-M4MR and KMB in terms of the average packet loss rate, with improvements of 1.76% and 26.94%, respectively. Its performance is similar to that of MADRL-MR and SCTF in this metric.

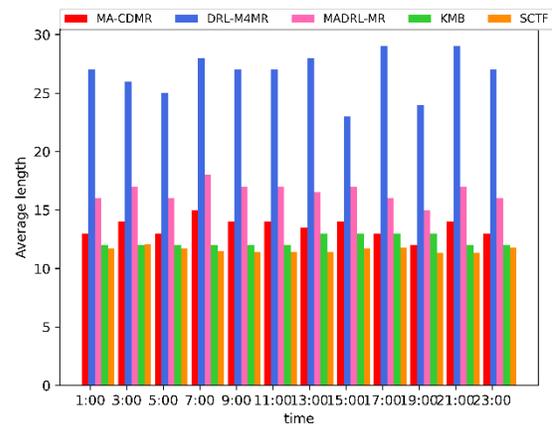

(a) Average multicast tree length

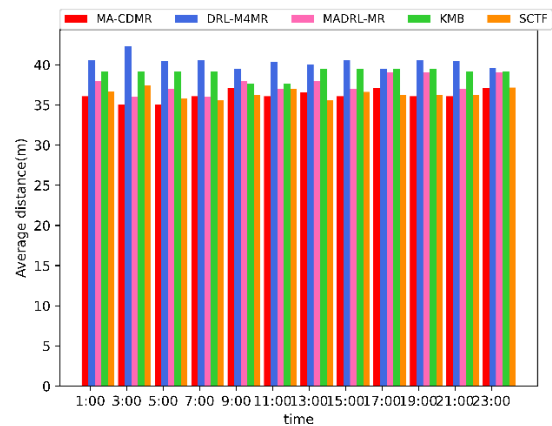

(b) Average link distance

Figure 23 Comparison of average tree length and link distance

Fig. 23(a) compares the average length of multicast trees constructed by MA-CDMR with those of DRL-M4MR, MADRL-MR, KMB, and SCTF. The results show that the multicast trees established by MA-CDMR have shorter lengths than those obtained with DRL-M4MR and MADRL-MR, which reflects the advantage of the MA-CDMR strategy in a multidomain environment. However, the tree lengths are longer than those obtained with KMB and SCTF, indicating that the algorithm considers more parameters and involves more nodes when selecting nodes for the multicast path. The MA-CDMR algorithm makes a trade-off when balancing length and performance.

Fig. 23(b) compares the average distance between wireless AP nodes in the multicast trees constructed by MA-CDMR with DRL-M4MR, MADRL-MR, KMB, and SCTF. The results indicate that MA-CDMR performs better than other methods in terms of the average distance between AP nodes in the multicast trees. Although the average length of the multicast trees in MA-CDMR is longer than that obtained with KMB and SCTF, as shown in Fig 23(a), the distances between the AP nodes do not exhibit the same trend. This suggests that the algorithm considers the distances between wireless AP nodes and achieves favorable results. The multiagent strategy of MA-CDMR for cross-domain multicasting outperforms the single-domain multiagent subtask strategy of MADRL-MR.

## VII. CONCLUSION

In this paper, the MA-CDMR algorithm, an intelligent cross-domain multicast routing method based on multiagent deep reinforcement learning in the SDWN multicontroller domain, is proposed. First, we analyze and model the cross-domain multicast routing problem in multicontroller domains, decomposing the multicast tree into interdomain multicast trees and multiple intradomain multicast trees. We construct these trees in two stages to approximate the optimal solution. Second, we design two types of agents: interdomain and intradomain. The interdomain agents are responsible for constructing interdomain multicast trees, whereas each intradomain agent is responsible for constructing the corresponding intradomain multicast tree. We design the state space of the agents via traffic data collected with the SDWN multicontroller domain network architecture. On the basis of the characteristics of the tasks for the two types of agents, we design corresponding action spaces. The action space of the interdomain agents is the set of edges between all domains, and each action corresponds to selecting an interdomain path to each neighboring domain. The action space of the intradomain agents is the set of intradomain nodes in the corresponding domain, and each action corresponds to selecting a node to add to the intradomain multicast tree. Finally, we design appropriate reward functions on the basis of different scenarios.

Extensive comparative experiments verify that the MA-CDMR algorithm outperforms classic multicast tree optimization methods such as KMB and SCTF, as well as the DRL-M4MR and MADRL-MR algorithms, which use reinforcement learning for multicast tree construction, in the context of multicontroller domains.

Furthermore, this work focuses on the control plane routing construction problem. In future research, exploring optimal practices for controller deployment would be beneficial. This includes determining the optimal controller positions, quantities, and layout to maximize multicast routing performance. Additionally, exploring P4 programming aspects of the data plane could further enhance multicast routing performance and flexibility. By optimizing the P4 programming model and designing efficient multicast routing functionalities, intelligent algorithms could be combined with P4 programming to achieve more intelligent, adaptive, and high-quality multicast routing.